\documentclass[10pt,aps,prd,twocolumn,showpacs,superscriptaddress,longbibliography,nofootinbib]{revtex4-2}
\usepackage{mathrsfs,amsmath,amsthm,latexsym,amssymb,amsfonts,epsfig,cancel,enumerate,graphicx,txfonts,diagbox}
\usepackage[utf8]{inputenc}
\usepackage[T1]{fontenc}
\usepackage{lmodern}
\usepackage{multirow}
\usepackage[colorlinks=true,linkcolor=blue,citecolor=blue,urlcolor=blue]{hyperref}
\usepackage{orcidlink}

\usepackage{xcolor}
\usepackage{booktabs}
\definecolor{navy}{RGB}{0,0,150}
\usepackage{appendix}

\allowdisplaybreaks

\newcommand{\RGU}{Department of Physics, The Assam Royal Global University, Guwahati-781035, Assam, India}

\newcommand{\UCC}{Programa de P\'os-Gradua\c c\~ao em F\'{\i}sica \& Coordena\c c\~ao do Curso de F\'{\i}sica -- Bacharelado, Universidade Federal do Maranh\~{a}o, 65085-580 S\~{a}o Lu\'{\i}s, Maranh\~{a}o, Brazil}

\begin{document}

\title{Thermodynamic Analysis of Charged AdS Black Holes with Cloud of Strings in \\[1ex] Einstein-Bumblebee Gravity via Tsallis Entropy }

\author{Faizuddin Ahmed\orcidlink{0000-0003-2196-9622}}
\email{faizuddinahmed15@gmail.com}
\affiliation{\RGU}

\author{Edilberto O. Silva\orcidlink{0000-0002-0297-5747}}
\email{edilberto.silva@ufma.br}
\affiliation{\UCC}

\date{\today}

\begin{abstract}
We investigate the thermodynamic properties of charged anti-de Sitter black holes surrounded by a cloud of strings in bumblebee gravity. In this framework, the cloud-of-strings parameter $\alpha$ and the Lorentz-violating parameter $\ell$ modify the horizon structure, the Hawking temperature, the free energies, the specific heat, and the critical behavior in the extended phase-space description. We derive the corresponding equation of state and show that the system exhibits a small--large black-hole phase transition of Van der Waals type. In particular, the critical quantities are deformed by both the cloud of strings and the bumblebee background, while the universal ratio is explicitly altered by Lorentz symmetry breaking. We also examine the Joule--Thomson expansion and analyze the associated inversion and isenthalpic curves, showing how the deformation parameters shift the boundary between heating and cooling regions. In addition, we extend the thermodynamic analysis to a Tsallis entropy-based framework and show that the non-extensive parameter $\delta$ significantly changes the temperature profile, stability windows, critical volume, free energies, and sparsity of Hawking radiation. Our results reveal that the combined effects of the string cloud, Lorentz violation, and non-extensive entropy lead to a substantially richer thermodynamic structure than that of the standard Reissner--Nordstr\"om--AdS black hole.
\end{abstract}

\maketitle

\section{Introduction}\label{sec:1}

General Relativity and the Standard Model of particle physics are two of the most successful theoretical frameworks describing the fundamental interactions in nature. General Relativity provides an accurate description of gravitational interactions, while the Standard Model successfully explains the electromagnetic, weak, and strong interactions. However, achieving a consistent unification of these two theories remains one of the most challenging problems in modern theoretical physics. Although several candidate theories of quantum gravity have been proposed, testing them experimentally or observationally typically requires access to the Planck energy scale ($\sim 10^{19}\,\mathrm{GeV}$), which is far beyond current experimental capabilities. Nevertheless, certain low-energy signals arising from underlying quantum gravity effects could still be observable. One important possibility is the spontaneous breaking of Lorentz symmetry, which has been widely studied as a potential low-energy signature of quantum gravity.

The thermodynamic description of black holes is one of the most remarkable bridges between gravitation, quantum theory, and statistical physics. The foundational works of Bekenstein, Bardeen, Carter, Hawking, and Hawking--Page established that black holes possess entropy, temperature, and a consistent set of mechanical and thermodynamic laws \cite{Bekenstein1973,BardeenCarterHawking1973,Hawking1975,HawkingPage1983,Davies1977}. These developments not only showed that horizons behave as genuine thermodynamic systems but also motivated the broader idea that black holes can serve as theoretical laboratories for exploring phase transitions, stability, and the microscopic aspects of gravity. Black hole thermodynamics is also closely related to quantum gravity, and quantum gravitational effects might become very important in the region near the horizon, where gravity is strong. Therefore, the thermodynamics of the bumblebee black holes might provide a new angle to probe Lorentz symmetry violation in the strong field region. In
addition, the local thermodynamic stability of the bumblebee black holes is an intriguing topic as well. It is usually characterized by heat capacity, which describes a thermodynamic system's ability to resist perturbations induced by exchanging a small amount of heat with its environment.

A major advance in this direction came with the formulation of black-hole thermodynamics in the extended phase space, where the cosmological constant is promoted to a thermodynamic pressure and the black-hole mass is interpreted as enthalpy \cite{KastorRayTraschen2009,Dolan2011,Dolan2011Compressibility,Kubiznak2012,Johnson2014,Kubiznak2017}. In this framework, AdS black holes exhibit a rich thermodynamic structure closely analogous to ordinary chemical systems, including Van der Waals-like criticality, first-order small--large black-hole phase transitions, universal critical behavior, coexistence curves, and heat-engine interpretations \cite{Kubiznak2012,Gunasekaran2012,MoLiu2014,MoXu2015,Johnson2014,Kubiznak2017}. Because of these results, extended phase-space thermodynamics has become one of the standard settings in which modifications of gravity or additional matter sectors can be tested quantitatively.

On the matter side, one particularly interesting deformation is provided by a cloud of strings. Letelier introduced this configuration as a gravitational source generated by one-dimensional extended objects, showing that it leads to exact solutions with a distinct contribution to the spacetime geometry \cite{Letelier1979,Letelier1983}. Since then, black holes surrounded by a cloud of strings have been studied in several contexts, including Lovelock gravity, regular black holes, and more recent analyses of their thermodynamic and geometric properties \cite{Ghosh2014PRD90,Ghosh2014PRD89,Rodrigues2022,Santos2025}. In the AdS setting, the cloud-of-strings parameter modifies the metric function in a simple but physically meaningful way, making it a useful phenomenological probe of how an external anisotropic matter distribution affects horizon thermodynamics and criticality \cite{GhaffarnejadYaraieFarsam2018,Santos2025}.

Another important source of deformation arises from Lorentz-symmetry violation in gravitational theories. The spontaneous breaking of Lorentz symmetry can be achieved by allowing the existence of a non-zero expectation value of some tensor field. A particularly well-studied framework is bumblebee gravity, in which a vector field acquires a nonzero vacuum expectation value and spontaneously breaks local Lorentz symmetry \cite{BluhmKostelecky2005,Bluhm2008}. This mechanism has motivated a broad literature on exact and approximate black-hole solutions, cosmological models, compact objects, and phenomenological signatures of Lorentz violation in the gravitational sector \cite{Casana2018,MalufNeves2021,FilhoEtAl2023,YangEtAl2023}. In the black-hole context, the bumblebee background can modify the radial metric component, the effective charge sector, the horizon structure, and the associated thermodynamic quantities. Therefore, when combined with a cloud of strings, it provides a natural arena in which two conceptually distinct deformations, one due to surrounding matter and the other due to spontaneous Lorentz breaking, act simultaneously on the black-hole system.

Motivated by these developments, in this paper, we investigate the thermodynamic properties of charged AdS black holes surrounded by a cloud of strings in bumblebee gravity. Our interest is twofold. First, we study how the cloud-of-strings parameter $\alpha$ and the Lorentz-violating parameter $\ell$ modify the standard thermodynamic observables, including the Hawking temperature, entropy, thermodynamic volume, Gibbs and Helmholtz free energies, specific heat, and the equation of state in the extended phase-space formulation. Second, we examine how these deformations affect the critical point and the small--large black-hole phase transition, with particular emphasis on the universal ratio $P_c v_c/T_c$, which is one of the clearest diagnostics of modified critical behavior in AdS black holes \cite{Kubiznak2012,Gunasekaran2012,MoLiu2014,MoXu2015}.

We also analyze the Joule--Thomson (JT) expansion in the present framework. In ordinary thermodynamics, the JT process characterizes the temperature response of a system undergoing an isenthalpic expansion. Its black-hole analogue has attracted considerable attention in the extended phase-space literature because it provides a complementary probe of the interplay between pressure, enthalpy, and phase structure \cite{Okcu2017}. Since the black-hole mass plays the role of enthalpy, the inversion curves and isenthalpic trajectories provide a useful way to identify heating and cooling regions and assess how background parameters reshape the thermal response. For the geometry considered here, it is therefore natural to ask how the string cloud and Lorentz-violating sector alter the inversion temperature and the associated JT landscape.

In addition to the standard Bekenstein--Hawking area law, it is of considerable interest to explore generalized entropy frameworks. Because gravity is intrinsically long-range and nonlocal, several authors have argued that non-extensive entropy functionals may capture aspects of black-hole thermodynamics that are not visible in the usual additive picture. The foundational proposal of Tsallis introduced a generalized non-additive entropy \cite{Tsallis1988}, and its application to black holes was later developed in the Tsallis--Cirto entropy proposal \cite{TsallisCirto2013}. More recently, Barrow proposed an entropy-area relation incorporating a fractal-like deformation of the horizon geometry, providing another route to generalized black-hole thermodynamics \cite{Barrow2020}. These approaches have motivated many studies because they can alter the temperature scaling, stability windows, and critical structure of black-hole systems. In the present work, we therefore extend the analysis to a Tsallis entropy-based description and discuss the corresponding Barrow-entropy interpretation. The thermodynamic properties of Schwarzschild-like black holes in bumblebee gravity, both with and without a cosmological constant, have been investigated in the literature (see Refs.~\cite{Ding2023a,Gomes2020,An2024,EslamPanah2025}). 

Finally, we also consider the sparsity of Hawking radiation. One of the notable results in the literature is that Hawking emission is not a continuous blackbody-like flux, but rather an extremely sparse cascade of quanta \cite{Page1976,GraySchusterVanBruntVisser2016}. This observation has motivated a series of investigations into how modified geometries, corrected entropies, and deformed black-hole backgrounds affect the temporal structure of the evaporation process. In our case, sparsity provides an additional diagnostic by which the combined effects of charge, AdS pressure, the string cloud, Lorentz violation, and non-extensive entropy can be assessed within a single framework.

The paper is organized as follows. In Sec.~\ref{sec:2}, we review the charged AdS black-hole solution with a cloud of strings in bumblebee gravity. In Sec.~\ref{sec:3}, we derive the standard thermodynamic quantities and analyze the associated critical behavior in the extended phase space. In Sec.~\ref{sec:4}, we study the Joule--Thomson expansion, including inversion and isenthalpic curves. In Sec.~\ref{sec:5}, we extend the discussion to the Tsallis entropy-based framework and examine the modifications in temperature, stability, free energies, criticality, and sparsity. We then comment on the Barrow-entropy interpretation and summarize our results in Sec.~\ref{sec:10}.

\section{Charged AdS BH with CS in Bumblebee Gravity: Background Metric}\label{sec:2}

A charged AdS black hole solution in the bumblebee gravity frame surrounded by a cloud of strings is characterized by the line element \cite{Li2026,Liu2025,Kala2026arxiv}
\begin{equation}
\mathrm{d} s^2=-f(r)\,dt^{2}+\frac{1+\ell}{f(r)}\,dr^{2}+r^{2}(d\theta ^{2}+\sin^{2}\theta\, d\phi^2),\label{metric}
\end{equation}
where the metric function $f(r)$ takes the form
\begin{align}
f(r)&=1-\alpha-\frac{2M}{r}+\frac{\sigma\,q^{2}}{r^{2}}-\frac{\Lambda_e}{3}\, r^2\,(1+\ell),\nonumber\\
\Lambda_e&=\lambda/\xi,\quad \sigma=\frac{1+\ell}{1+\ell/2}. \label{function}
\end{align}

Here, $\xi$ is the nonminimal coupling constant between gravity and the bumblebee field, with $b_{\mu}b^{\mu}=b^{2}=\text{const.}$, and $\ell=\xi b^2$ is the Lorentz-violating (LV) parameter. The quantity $\Lambda_e=\lambda/\xi$ plays the role of an effective cosmological constant, where $\lambda$ is a Lagrange multiplier enforcing the bumblebee field constraint \cite{Bluhm2008}. Finally, $\alpha$ is a dimensionless constant associated with the string-cloud matter distribution \cite{Letelier1979}, and $M$ and $q$ denote the geometric mass and electric charge of the black hole, respectively.

Several important limits can be identified directly from the metric function~(\ref{function}):
\begin{itemize}
    \item When $\ell=0$, the factor $\sigma\to 1$ and $\Lambda_e \to \Lambda$, so the space-time reduces to the Reissner--Nordstr\"om--AdS black hole solution surrounded by a cloud of strings.
    \item When both $\ell=0$ and $q=0$, the solution degenerates into the Letelier--AdS space-time \cite{Letelier1979}.
    \item When $q=0$ (but $\ell\neq 0$), the metric reduces to the Schwarzschild-like AdS black hole in bumblebee gravity, as derived in Ref.~\cite{Casana2018}.
\end{itemize}
These limiting cases confirm the solution's consistency and its connections to well-established spacetimes in the literature.

It is worth noting that the non-diagonal structure introduced by the bumblebee field modifies the effective charge coupling through $\sigma=(1+\ell)/(1+\ell/2)$. This factor satisfies $\sigma\to 1$ as $\ell\to 0$ and $\sigma>1$ for $\ell>0$, meaning that Lorentz violation effectively enhances the contribution of the electric charge to the gravitational potential. Similarly, the effective cosmological constant $\Lambda_e$ is rescaled relative to its GR counterpart, encoding the backreaction of the bumblebee field on the background geometry.

\section{Thermodynamics}\label{sec:3}

In this section, we investigate the thermodynamic properties of the black hole solution described above. We derive the Hawking temperature, entropy, thermodynamic volume, Gibbs and Helmholtz free energies, and specific heat capacity, working throughout within the extended phase-space framework.

A black hole is not only a gravitational system but also a special thermodynamic system, as its surface gravity $\kappa$ and horizon area $A$ closely resemble the temperature $T$ and entropy $S$ of ordinary thermodynamic systems, with the identifications \(T = \frac{\kappa}{2\pi},\quad S = \frac{A}{4}.\) The four laws of black hole thermodynamics were established in Ref.~\cite{Bardeen1973}. Modifications to the black hole geometry arising from Lorentz violation (LV) and Cloud of strings (CS) terms can alter these thermodynamic properties while still satisfying the generalized thermodynamic laws.

\subsection{Hawking Temperature and Entropy}

The Hawking temperature is the temperature associated with the quantum radiation emitted by a black hole, indicating that black holes behave like thermodynamic objects. 

The geometric mass parameter $M$ is determined by the condition that $f(r_h)=0$, where $r_h$ denotes the event horizon radius. From Eq.~(\ref{function}), this yields
\begin{equation}
    M=\frac{r_h}{2}\left[1-\alpha+\frac{\sigma\,q^{2}}{r_h^{2}}-\frac{\Lambda_e(1+\ell)}{3}\, r_h^2\right].\label{mass1}
\end{equation}

In the extended phase space framework, the effective cosmological constant is identified with the thermodynamic pressure $P$ via \cite{Kubiznak2012}
\begin{equation}
    P = -\frac{\Lambda_e(1+\ell)}{8\pi},\label{pressure}
\end{equation}
so that $-\Lambda_e(1+\ell)/3 = 8\pi P/3$. Substituting into Eq.~(\ref{mass1}), the geometric mass reads
\begin{equation}
    M=\frac{r_h}{2}\left[1-\alpha+\frac{\sigma\,q^{2}}{r_h^{2}}+\frac{8\pi P}{3}\, r_h^2\right].\label{mass3}
\end{equation}
Note that in this framework, the pressure $P$ is a thermodynamic variable independent of the solution's other parameters.
\begin{figure*}[tbhp]
\centering
\includegraphics[width=0.98\textwidth]{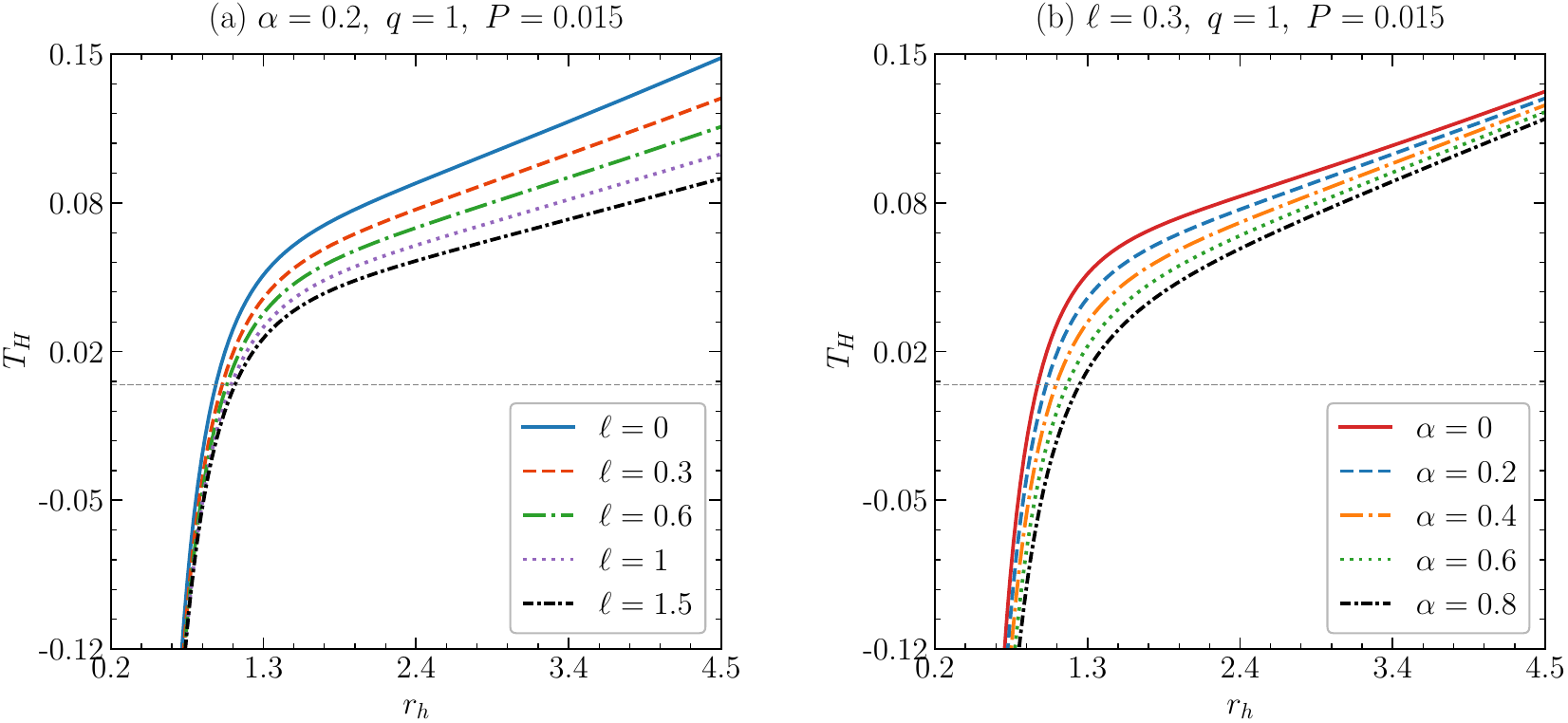}
\caption{Hawking temperature $T_H$ as a function of the horizon radius $r_h$. Panel (a) shows the effect of the Lorentz-violating parameter $\ell$ for fixed $\alpha=0.2$, $q=1$, and $P=0.015$. Panel (b) shows the effect of the cloud-of-strings parameter $\alpha$ for fixed $\ell=0.3$, $q=1$, and $P=0.015$. The curves show that increasing $\ell$ suppresses the temperature through the global factor $(1+\ell)^{-1/2}$, while increasing $\alpha$ lowers the overall temperature scale by reducing the effective contribution of the $(1-\alpha)$ term.}
\label{fig:temperature}
\end{figure*}

The Hawking temperature is derived from the surface gravity $\kappa$ evaluated at the event horizon. For a static metric of the form~(\ref{metric}), the surface gravity is \cite{Hawking1975,Bekestein1972,Bekestein1973,Bardeen1973}
\begin{align}
    \kappa &= -\frac{1}{2}\lim_{r \to r_h}\frac{\partial_r g_{tt}}{\sqrt{-g_{tt}\,g_{rr}}},
\end{align}
where the factor $\sqrt{-g_{tt}\,g_{rr}}=\sqrt{f(r)\cdot(1+\ell)/f(r)}=\sqrt{1+\ell}$ is constant in $r$, so that
\begin{equation}
    \kappa = \frac{f'(r_h)}{2\sqrt{1+\ell}}.\label{surfcae-gravity}
\end{equation}
Computing $f'(r)=2M/r^2 - 2\sigma q^2/r^3 - 2\Lambda_e(1+\ell)r/3$ and evaluating at $r=r_h$ using Eq.~(\ref{mass3}), one finds
\begin{equation}
    \kappa = \frac{1-\alpha-\dfrac{\sigma q^{2}}{r_h^{2}}+8\pi P\,r_h^2}{2 r_h \sqrt{1+\ell}}.\label{kappa}
\end{equation}
The Hawking temperature $T_H=\kappa/(2\pi)$ is therefore
\begin{equation}
    T_H=\frac{1-\alpha-\dfrac{\sigma q^{2}}{r_h^{2}}+8\pi P\,r_h^2}{4\pi r_h \sqrt{1+\ell}}.\label{temperature}
\end{equation}
One can verify the limiting cases: setting $\ell=0$ and $\alpha=0$ recovers the standard RN-AdS Hawking temperature, while taking $q=0$ additionally gives the Schwarzschild-AdS result. It is also clear from~(\ref{temperature}) that the LV parameter $\ell$ suppresses the temperature globally through the factor $(1+\ell)^{-1/2}$ in the denominator, while the string-cloud parameter $\alpha$ reduces the effective mass contribution.

The ADM mass of the thermodynamic system, which accounts for the normalization induced by the bumblebee field, is given by \cite{AbbottDeser1982}
\begin{equation}
    \mathcal{M}=\frac{M}{\sqrt{1+\ell}}.\label{adm}
\end{equation}

The entropy of the system can also be obtained from the first law of thermodynamics, $d\mathcal{M}=T_H\,dS$, which gives \cite{Bardeen1973,Hawking1983}
\begin{equation}
    S=\int \frac{d\mathcal{M}}{T_H}.
\end{equation}
Differentiating Eq.~(\ref{mass3}) with respect to $r_h$ (at fixed $P$, $q$, $\alpha$) and using~(\ref{temperature}), a direct computation yields
\begin{equation}
    S=\pi r_h^2,\label{entropy}
\end{equation}
which is the standard Bekenstein-Hawking area law, $S=A/4$, with $A=4\pi r_h^2$. Remarkably, the entropy retains its standard form despite the modifications introduced by both the cloud of strings and the bumblebee field; these modifications affect the temperature but not the area--entropy relation itself.

The dependence of the temperature on $r_h$ is illustrated in Fig.~\ref{fig:temperature}. The two panels separately display the influence of the Lorentz-violating parameter $\ell$ and the string-cloud parameter $\alpha$. In both cases, the behavior at small horizon radius is dominated by the electric term, while for sufficiently large $r_h$ the AdS contribution proportional to $P r_h^2$ makes the temperature grow. This produces the familiar non-monotonic profile associated with a thermodynamic competition between charge and AdS pressure.

Figure~\ref{fig:temperature}(a) makes the effect of Lorentz violation particularly transparent: larger values of $\ell$ shift the whole temperature profile downward, indicating that the bumblebee background cools the black hole at fixed horizon radius. By contrast, Fig.~\ref{fig:temperature}(b) shows that increasing the string-cloud density parameter $\alpha$ also reduces the temperature, but through a different mechanism, namely the weakening of the constant contribution to the numerator of Eq.~(\ref{temperature}). In both panels, the existence of a minimum suggests the possibility of multiple thermodynamic branches for the same temperature.

\subsection{Free Energies and Specific Heat}

Expressing $r_h=\sqrt{S/\pi}$ and substituting into Eq.~(\ref{mass3}), the ADM mass in terms of entropy reads
\begin{equation}
    \mathcal{M}=\frac{1}{2\sqrt{1+\ell}}\sqrt{\frac{S}{\pi}}\left[1-\alpha+\frac{\sigma \pi q^2}{S}+\frac{8P}{3}\,S\right].\label{mass4}
\end{equation}
Similarly, the Hawking temperature and thermodynamic volume become
\begin{align}
    T_H &= \frac{1-\alpha-\dfrac{\sigma \pi q^{2}}{S}+8 P S}{4\sqrt{1+\ell}\sqrt{\pi S}},\label{temperature2}\\[6pt]
    V &= \left(\frac{\partial \mathcal{M}}{\partial P}\right)_{S,q,\alpha}=\frac{4\pi r_h^3}{3\sqrt{1+\ell}}.\label{volume}
\end{align}
The thermodynamic volume $V$ differs from the naive geometric volume $V_{\rm geo}=4\pi r_h^3/3$ by the factor $(1+\ell)^{-1/2}$, which is a direct signature of the Lorentz-violating bumblebee background. As $\ell$ increases, the thermodynamic volume decreases relative to the geometric one.

The Gibbs free energy $G=\mathcal{M}-T_H S$ \cite{Hawking1983} controls the thermodynamic stability of the black hole at fixed $T_H$ and $P$. Using Eqs.~(\ref{mass4}) and~(\ref{temperature2}),
\begin{equation}
    G=\frac{\sqrt{S/\pi}}{4\sqrt{1+\ell}}\left(1-\alpha+\frac{3\sigma \pi q^2}{S}-\frac{8P}{3}S\right).\label{gibbs}
\end{equation}
The sign of $G$ is particularly important: a negative Gibbs free energy signals that the black hole phase is thermodynamically preferred over the thermal radiation (Hawking--Page transition). The pressure-dependent term $-8PS/3$ makes $G$ decrease with increasing $P$, favoring the black hole phase at high pressures, consistent with the general behavior of AdS black holes.

The Helmholtz free energy $F=\mathcal{M}-PV$ \cite{Hawking1983} of the black hole system is obtained as 
\begin{align}
    F&=\frac{r_h}{2\sqrt{1+\ell}}\left[1-\alpha+\frac{\sigma q^2}{r_h^2}\right]\nonumber\\
    &=\frac{1}{2\sqrt{(1+\ell)\pi}}\left[(1-\alpha) S^{1/2}+\sigma \pi q^2 S^{-1/2}\right],\label{free}
\end{align}
which is manifestly positive and independent of $P$, reflecting its role as the thermodynamic potential at fixed volume.

The specific heat at constant pressure is a key indicator of local thermodynamic stability. It is given by \cite{Hawking1975}
\begin{equation}
    C_p = T_H\left(\frac{dS}{dT_H}\right)_P.
\end{equation}
To compute this, we first note that $S=\pi r_h^2$, so $dS=2\pi r_h\,dr_h$. Differentiating $T_H$ in Eq.~(\ref{temperature}) with respect to $r_h$ at fixed $P$ gives
\begin{equation}
\frac{dT_H}{dr_h}=\frac{-1+\alpha+\dfrac{3\sigma q^2}{r_h^2}+8\pi P r_h^2}{4\pi r_h^2\sqrt{1+\ell}},
\end{equation}
so that $dS/dT_H = 2\pi r_h\,(dT_H/dr_h)^{-1}$. Combining with~(\ref{temperature}), the specific heat is
\begin{align}
    C_p = 2\pi r_h^2\,\frac{1-\alpha-\dfrac{\sigma q^2}{r_h^2}+8\pi P r_h^2}{-1+\alpha+\dfrac{3\sigma q^2}{r_h^2}+8\pi P r_h^2}.\label{heat-1}
\end{align}
The sign of $C_p$ determines local thermodynamic stability: $C_p>0$ corresponds to a locally stable black hole, while $C_p<0$ indicates an unstable configuration prone to evaporation. The numerator of~(\ref{heat-1}) is proportional to $T_H$ (by comparison with Eq.~(\ref{temperature})), so $C_p$ changes sign only when the denominator vanishes, i.e., at
\begin{equation}
    -1+\alpha+\frac{3\sigma q^2}{r_h^2}+8\pi P r_h^2 = 0,
\end{equation}
which defines the radius $r_h^*$ at which a second-order phase transition (Davies point) occurs. At this point, $C_p$ diverges, signaling a change in stability.

\begin{figure*}[t]
\centering
\includegraphics[width=0.98\textwidth]{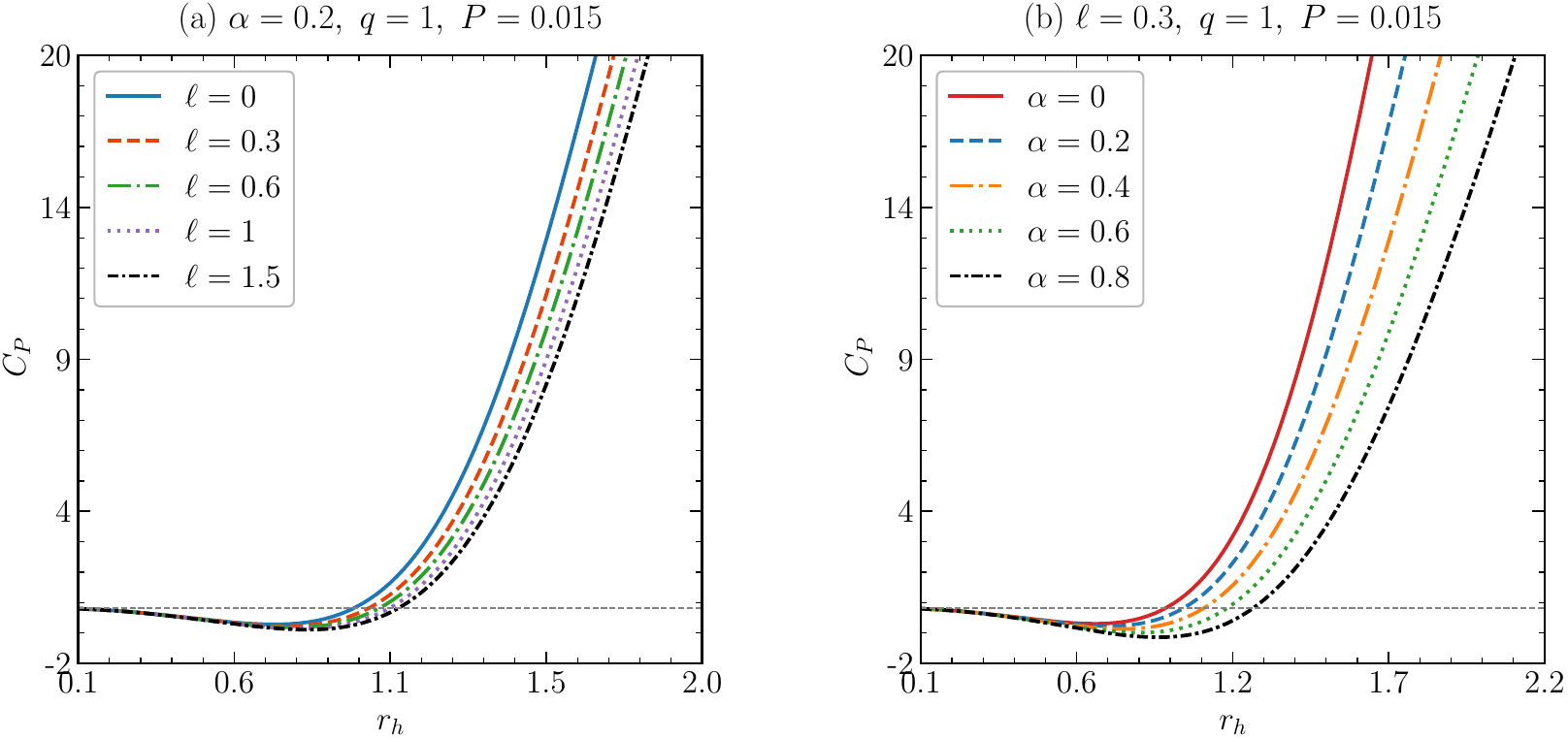}
\caption{Specific heat at constant pressure $C_P$ as a function of $r_h$. Panel (a) shows the dependence on the Lorentz-violating parameter $\ell$ for fixed $\alpha=0.2$, $q=1$, and $P=0.015$, while panel (b) shows the dependence on $\alpha$ for fixed $\ell=0.3$, $q=1$, and $P=0.015$. The vertical divergences correspond to Davies points, where the denominator of Eq.~(\ref{heat-1}) vanishes and the black hole changes from a locally unstable branch ($C_P<0$) to a locally stable branch ($C_P>0$).}
\label{fig:specificheat}
\end{figure*}

The stability structure is displayed in Fig.~\ref{fig:specificheat}. The divergence of $C_p$ separates branches with negative and positive specific heat, identifying the transition between unstable and locally stable configurations. The location of this divergence depends sensitively on both $\ell$ and $\alpha$.

From Fig.~\ref{fig:specificheat}(a), one sees that Lorentz violation shifts the divergence structure and therefore modifies the location of the stability threshold. Meanwhile, Fig.~\ref{fig:specificheat}(b) shows that the string cloud has a comparable thermodynamic effect: as $\alpha$ increases, the divergence moves and the width of the unstable branch changes. These results confirm that both deformations alter the local thermodynamic response in a nontrivial way.

\begin{figure*}[t]
\centering
\includegraphics[width=0.98\textwidth]{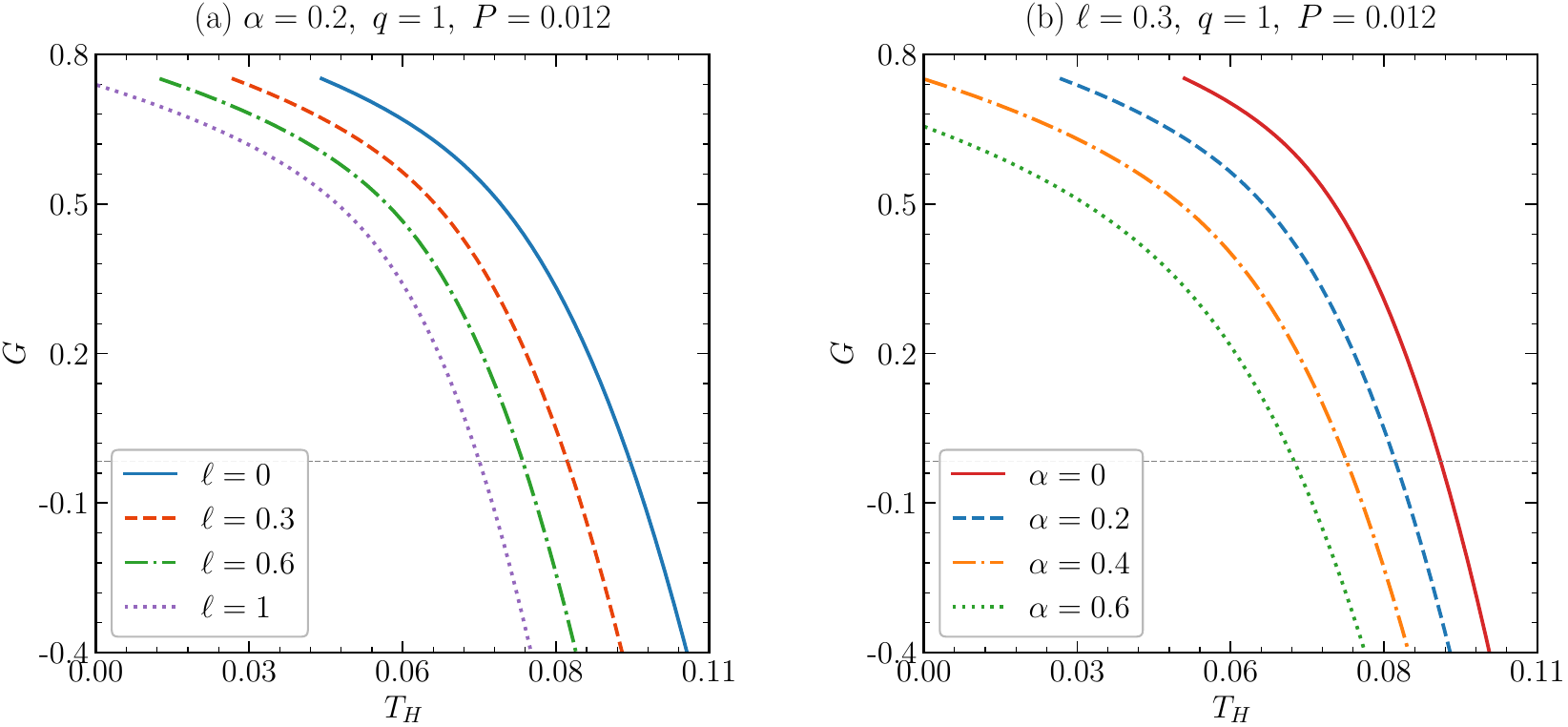}
\caption{Gibbs free energy $G$ as a function of Hawking temperature $T_H$. Panel (a) shows the effect of $\ell$ for fixed $\alpha=0.2$, $q=1$, and $P=0.012$, whereas panel (b) shows the effect of $\alpha$ for fixed $\ell=0.3$, $q=1$, and $P=0.012$. The swallowtail-like behavior signals the presence of competing thermodynamic branches and is the standard signature of a first-order small--large black-hole phase transition in the extended phase-space picture.}
\label{fig:gibbs}
\end{figure*}

The global thermodynamic behavior is encoded in the Gibbs free energy shown in Fig.~\ref{fig:gibbs}. The multibranch structure is characteristic of AdS black holes with first-order phase transitions, and the appearance of the swallowtail reflects the coexistence of small and large black-hole phases.

As seen in Fig.~\ref{fig:gibbs}, both $\ell$ and $\alpha$ deform the size and position of the swallowtail. This means that Lorentz symmetry breaking and the cloud of strings affect not only local stability, through $C_P$, but also global phase preference, through $G$. The branch with lower Gibbs free energy is thermodynamically favored, and the intersection structure among branches identifies the transition temperature.

\subsection{Modified First Law of Thermodynamics}

Since the ADM mass $\mathcal{M}$ in Eq.~(\ref{mass4}) depends on the entropy $S$, charge $q$, pressure $P$, and string-cloud parameter $\alpha$, the differential of $\mathcal{M}$ takes the form
\begin{equation}
    d\mathcal{M}=T\,dS+\Phi_q\,dq+V\,dP+\Phi_{\alpha}\,d\alpha,\label{differential}
\end{equation}
where the conjugate thermodynamic potentials are identified as
\begin{align}
T &= \left(\frac{\partial \mathcal{M}}{\partial S}\right)_{P,q,\alpha}
  = \frac{1-\alpha-\dfrac{\sigma\pi q^2}{S}+8PS}{4\sqrt{\pi S(1+\ell)}},\label{temp-first-law}\\[6pt]
\Phi_q &= \left(\frac{\partial \mathcal{M}}{\partial q}\right)_{P,S,\alpha}
       = \frac{\sigma q}{\sqrt{1+\ell}}\sqrt{\frac{\pi}{S}}
       = \frac{\sigma q}{r_h\sqrt{1+\ell}},\label{potential-1}\\[6pt]
\Phi_{\alpha} &= \left(\frac{\partial \mathcal{M}}{\partial \alpha}\right)_{P,S,q}
              = -\frac{1}{2\sqrt{1+\ell}}\sqrt{\frac{S}{\pi}}
              = -\frac{r_h}{2\sqrt{1+\ell}},\label{potential-2}\\[6pt]
V &= \left(\frac{\partial \mathcal{M}}{\partial P}\right)_{S,q,\alpha}
  = \frac{4\pi r_h^3}{3\sqrt{1+\ell}}.\label{potential-3}
\end{align}
One confirms that Eq.~(\ref{temp-first-law}) reproduces exactly the Hawking temperature~(\ref{temperature2}), as required by thermodynamic consistency.

A few remarks on the physical content of these potentials are in order. The electric potential $\Phi_q$ is enhanced relative to the standard RN-AdS result by the factor $\sigma/(1+\ell)^{1/2}=(1+\ell)^{1/2}/(1+\ell/2)$, which exceeds unity for all $\ell>0$. Thus, Lorentz violation effectively strengthens the electrostatic coupling. The string-cloud potential $\Phi_\alpha$ is negative, reflecting the fact that increasing $\alpha$ (i.e., increasing the string density) lowers the total energy of the system, consistent with the role of $\alpha$ in reducing the effective gravitational mass through the term $(1-\alpha)$ in~(\ref{function}).

\textbf{Generalized Smarr relation.} Applying Euler's theorem for homogeneous functions to the mass $\mathcal{M}(S,q,P,\alpha)$ --- with scaling dimensions $[\mathcal{M}]\sim L$, $[S]\sim L^2$, $[q]\sim L$, $[P]\sim L^{-2}$, $[\alpha]\sim L^0$ --- one obtains the generalized Smarr relation
\begin{equation}
\mathcal{M} = 2(T S - P V) + \Phi_{q}\, q + 2\,\Phi_\alpha\,\alpha.\label{smarr-wrong}
\end{equation}
However, since $\alpha$ is dimensionless (scaling dimension zero), its contribution to the Euler relation vanishes, and the modified Smarr relation simplifies to
\begin{equation}
\mathcal{M} = 2\,T S - 2\,P V + \Phi_{q}\, q.
\label{smarr}
\end{equation}
To verify this, we substitute the explicit expressions for $T$, $V$, and $\Phi_q$ from Eqs.~(\ref{temp-first-law})--(\ref{potential-3}):
\begin{align}
2TS &= 2 \cdot \frac{1-\alpha-\frac{\sigma\pi q^2}{S}+8PS}{4\sqrt{\pi S(1+\ell)}} \cdot S \nonumber\\
    &= \frac{\sqrt{S/\pi}}{2\sqrt{1+\ell}}\left(1-\alpha-\frac{\sigma\pi q^2}{S}+8PS\right),\\
2PV &= \frac{8\pi P r_h^3}{3\sqrt{1+\ell}}=\frac{8P}{3\sqrt{1+\ell}}\left(\frac{S}{\pi}\right)^{3/2},\\
\Phi_q q &= \frac{\sigma q^2}{r_h\sqrt{1+\ell}}=\frac{\sigma q^2\sqrt{\pi}}{\sqrt{S(1+\ell)}}.
\end{align}
Combining these three terms and simplifying yields exactly $\mathcal{M}$ as given in~(\ref{mass4}), confirming that the Smarr relation~(\ref{smarr}) is satisfied identically. This self-consistency provides a non-trivial check of all thermodynamic quantities derived above.

\subsection{Thermodynamic Criticality}

To study small--large black hole phase transitions, we derive the equation of state by solving Eq.~(\ref{temperature}) for $P$:
\begin{equation}
    P=\frac{\sqrt{1+\ell}\, T_H}{2 r_h} - \frac{1-\alpha}{8 \pi r_h^2} + \frac{\sigma q^2}{8 \pi r_h^4}.\label{critical-1}
\end{equation}
This has the same functional structure as the Van der Waals equation of state when the horizon radius $r_h$ is identified with the specific volume $v=2r_h$ \cite{Kubiznak2012}, yielding
\begin{equation}
    P=\sqrt{1+\ell}\,\frac{T_H}{v}-\frac{c_1}{v^2}+\frac{c_2}{v^4},\label{critical-3}
\end{equation}
where we have defined the shorthand coefficients
\begin{equation}
    c_1=\frac{1-\alpha}{2\pi},\quad c_2=\frac{2\sigma q^2}{\pi}=\frac{8q^2(1+\ell)}{\pi(2+\ell)}.\label{critical-4}
\end{equation}
The analogy with the Van der Waals gas $P=T/(v-b)-a/v^2$ is evident: the first term plays the role of the thermal pressure, the second is an attractive correction (reduced by the string cloud through $1-\alpha$), and the third arises from the electric charge and replaces the excluded-volume repulsion.

The critical point $(v_c, T_c, P_c)$ is determined by the simultaneous conditions of an inflection point in the $P$--$v$ isotherm:
\begin{equation}
    \left(\frac{\partial P}{\partial v}\right)_{T_H}=0,\qquad \left(\frac{\partial^2 P}{\partial v^2}\right)_{T_H}=0.\label{critical-5}
\end{equation}
Differentiating Eq.~(\ref{critical-3}),
\begin{align}
\frac{\partial P}{\partial v}\bigg|_{T_H} &= -\frac{\sqrt{1+\ell}\,T_H}{v^2}+\frac{2c_1}{v^3}-\frac{4c_2}{v^5}=0,\label{dPdv}\\
\frac{\partial^2 P}{\partial v^2}\bigg|_{T_H} &= \frac{2\sqrt{1+\ell}\,T_H}{v^3}-\frac{6c_1}{v^4}+\frac{20c_2}{v^6}=0.\label{d2Pdv2}
\end{align}
Eliminating $T_H$ between these two equations by multiplying~(\ref{dPdv}) by $2/v$ and subtracting from~(\ref{d2Pdv2}), one obtains
\begin{equation}
    -\frac{2c_1}{v_c^4}+\frac{12c_2}{v_c^6}=0 \implies v_c^2 = \frac{6c_2}{c_1},
\end{equation}
which gives the critical specific volume:
\begin{equation}
v_c = \sqrt{\frac{6c_2}{c_1}}=\sqrt{\frac{1+\ell}{1+\ell/2}\,\frac{24}{1-\alpha}}\,q.\label{vc}
\end{equation}
Substituting $v_c$ back into Eq.~(\ref{dPdv}) to solve for $T_c$:
\begin{align}
T_c &= \frac{2c_1 v_c^{-1} - 4c_2 v_c^{-3}}{\sqrt{1+\ell}}
      = \frac{2c_1}{v_c\sqrt{1+\ell}}\left(1-\frac{2c_2}{c_1 v_c^2}\right)\nonumber\\
    &= \frac{2c_1}{v_c\sqrt{1+\ell}}\left(1-\frac{1}{3}\right)
     = \frac{4c_1}{3v_c\sqrt{1+\ell}},
\end{align}
where we used $v_c^2=6c_2/c_1$ in the last step. Inserting the explicit form of $v_c$:
\begin{equation}
T_c = \frac{4c_1^{3/2}}{3\sqrt{6(1+\ell)\,c_2}}=\frac{(1-\alpha)^{3/2}(1+\ell/2)^{1/2}}{3\sqrt{6}\,\pi q\,(1+\ell)}.\label{Tc}
\end{equation}
The critical pressure follows by substituting $(v_c, T_c)$ into~(\ref{critical-3}):
\begin{align}
P_c &= \sqrt{1+\ell}\,\frac{T_c}{v_c}-\frac{c_1}{v_c^2}+\frac{c_2}{v_c^4}\nonumber\\
    &= \frac{4c_1}{3v_c^2}-\frac{c_1}{v_c^2}+\frac{c_1}{6v_c^2}=\frac{c_1}{v_c^2}\left(\frac{4}{3}-1+\frac{1}{6}\right)\nonumber\\
    &= \frac{c_1^2}{12c_2}=\frac{(1-\alpha)^2(1+\ell/2)}{96\pi q^2(1+\ell)}.\label{Pc}
\end{align}

The universal critical ratio is
\begin{equation}
    \rho_c=\frac{P_c\,v_c}{T_c}=\frac{3\sqrt{1+\ell}}{8}.\label{critical-7}
\end{equation}
This is a remarkable result: while in the standard Van der Waals gas and in RN-AdS black holes $\rho_c=3/8$, here $\rho_c$ depends explicitly on the Lorentz-violating parameter $\ell$. For $\ell>0$, we have $\rho_c>3/8$, meaning the critical ratio is enhanced by Lorentz violation. This deviation indicates that the universality class of the small--large black hole phase transition is modified by the bumblebee field. Crucially, $\rho_c$ is independent of both $\alpha$ and $q$, so the departure from the standard value $3/8$ is a clean, unambiguous signature of Lorentz symmetry breaking.

\begin{figure*}[t]
\centering
\includegraphics[width=0.98\textwidth]{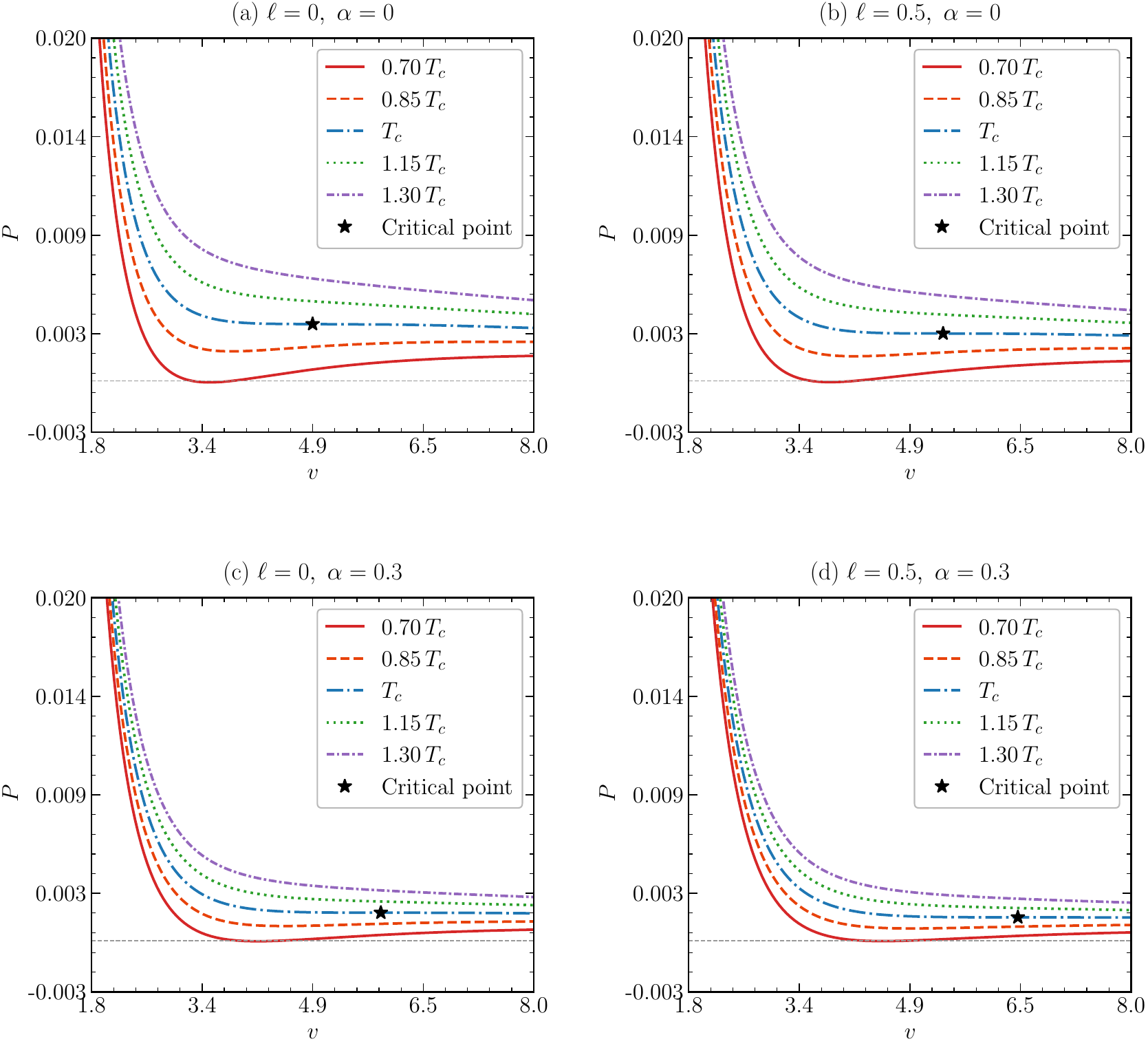}
\caption{$P$--$v$ isotherms for four representative choices of the deformation parameters: (a) $\ell=0$, $\alpha=0$; (b) $\ell=0.5$, $\alpha=0$; (c) $\ell=0$, $\alpha=0.3$; and (d) $\ell=0.5$, $\alpha=0.3$, with $q=1$ in all panels. Each panel includes isotherms below, at, and above the critical temperature, together with the critical point marked by a star. The subcritical curves show the Van der Waals-like oscillatory behavior associated with a small--large black-hole phase transition.}
\label{fig:Pv}
\end{figure*}

The phase structure encoded in the equation of state is displayed in Fig.~\ref{fig:Pv}. The isotherms below the critical temperature develop an oscillatory region characteristic of a first-order transition, while the critical isotherm marks the endpoint of the coexistence line.

Figure~\ref{fig:Pv} confirms that the present black hole exhibits the same qualitative thermodynamic pattern as a real fluid, but with the quantitative location of the critical point shifted by $\ell$ and $\alpha$. Increasing $\ell$ and $\alpha$ changes both the position of the inflection point and the shape of the isotherms, indicating that Lorentz violation and the cloud of strings deform the effective attractive and repulsive sectors of the equation of state.

\begin{figure*}[t]
\centering
\includegraphics[width=0.98\textwidth]{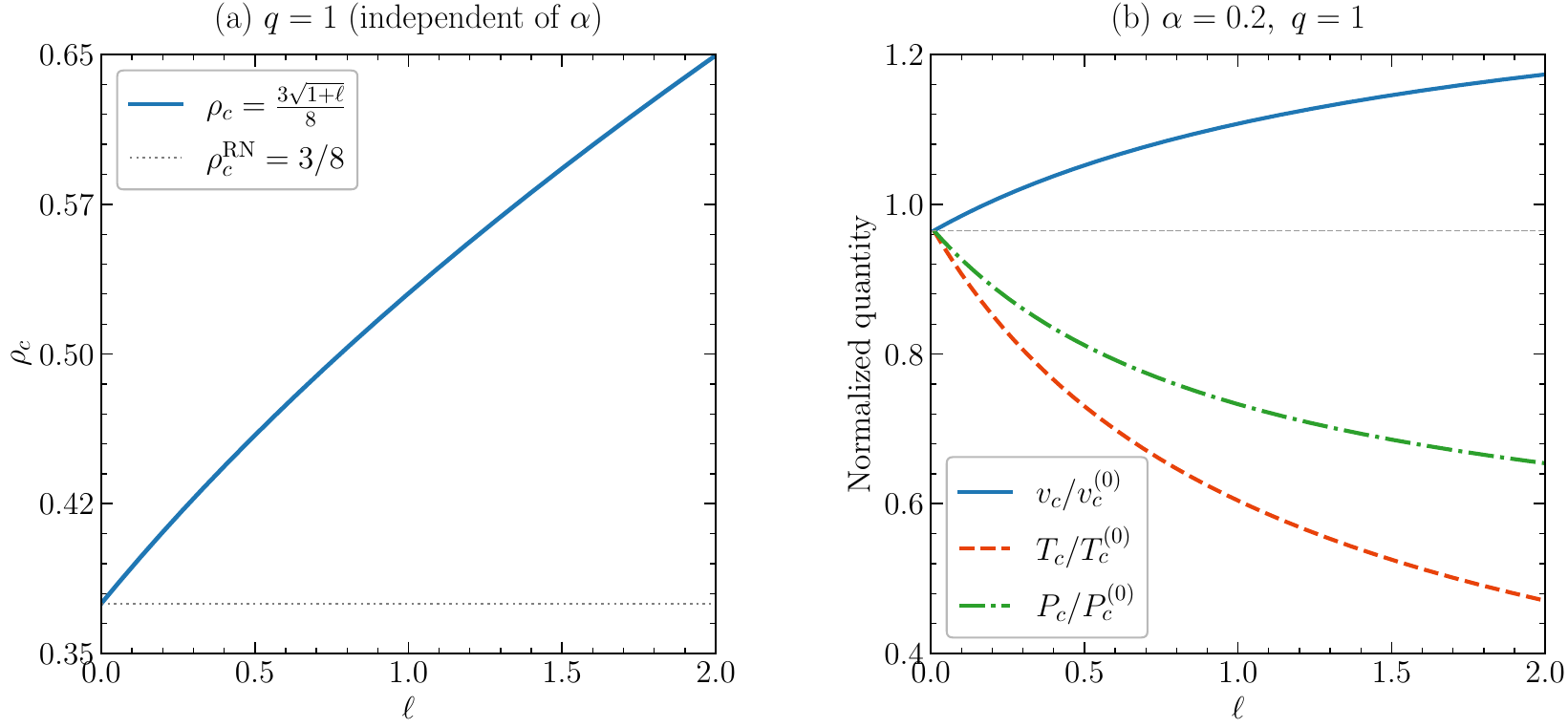}
\caption{Critical quantities in the standard thermodynamic framework. Panel (a) shows the universal critical ratio $\rho_c=P_c v_c/T_c=3\sqrt{1+\ell}/8$ as a function of the Lorentz-violating parameter $\ell$ for $q=1$. Since $\rho_c$ is independent of the cloud-of-strings parameter $\alpha$, all curves corresponding to different $\alpha$ coincide exactly; for this reason, only the single resulting profile is displayed. The horizontal dotted line indicates the standard RN--AdS value $\rho_c=3/8$. Panel (b) shows the normalized critical quantities $v_c/v_c^{(0)}$, $T_c/T_c^{(0)}$, and $P_c/P_c^{(0)}$ as functions of $\ell$ for fixed $\alpha=0.2$ and $q=1$.}
\label{fig:criticality}
\end{figure*}

These deformations become even clearer in Fig.~\ref{fig:criticality}, where we directly plot the universal ratio $\rho_c$ and the normalized critical quantities. Since $\rho_c$ depends only on $\ell$, this figure provides a particularly clean graphical signature of Lorentz violation in the critical sector.

Figure~\ref{fig:criticality}(a) makes explicit that the universal critical ratio is independent of the cloud-of-strings parameter $\alpha$ and is controlled exclusively by the Lorentz-violating parameter $\ell$. This is one of the most distinctive results of the present model, since it isolates the effect of Lorentz violation in a single thermodynamic observable. Meanwhile, Fig.~\ref{fig:criticality}(b) shows that the critical volume, temperature, and pressure are affected differently by Lorentz violation: the critical volume increases, whereas the critical temperature and critical pressure decrease in normalized units. This split behavior highlights the nontrivial way in which the bumblebee background reshapes the location of the critical point.

\section{Joule-Thomson Expansion}\label{sec:4}

The Joule-Thomson (JT) expansion is an isenthalpic process in which a gas (or, by analogy, a black hole) expands from high pressure to low pressure through a porous plug, with no work done on the surroundings and no heat exchanged. For black holes in the extended phase space, the mass $\mathcal{M}$ plays the role of the enthalpy \cite{Okcu2017}, so the JT process is one in which $\mathcal{M}$ remains fixed while $P$ decreases.

\subsection{Joule-Thomson Coefficient}

The key quantity in a JT process is the Joule-Thomson coefficient $\mu_{JT}$, defined as the rate of change of temperature with pressure at fixed enthalpy (mass):
\begin{equation}
    \mu_{JT}=\left(\frac{\partial T_H}{\partial P}\right)_{\mathcal{M}}.\label{JT-1}
\end{equation}
Using the cyclic identity for partial derivatives,
\begin{equation}
    \mu_{JT}=-\frac{1}{C_p}\left(\frac{\partial \mathcal{M}}{\partial P}\right)_{T_H}=-\frac{V}{C_p}\left(T_H\,\frac{\partial \ln V}{\partial T_H}\bigg|_P - 1\right).
\end{equation}
A more direct route uses the triple product rule applied to $T_H(P, r_h)$ and $\mathcal{M}(P, r_h)$, giving
\begin{equation}
    \mu_{JT}=\left(\frac{\partial T_H}{\partial P}\right)_{\mathcal{M}}=\frac{1}{C_p}\left[T_H\left(\frac{\partial V}{\partial T_H}\right)_P - V\right].\label{JT-coeff}
\end{equation}

To evaluate~(\ref{JT-coeff}) explicitly for our black hole, it is convenient to work at fixed $r_h$ and vary $P$. From the temperature~(\ref{temperature}) and volume~(\ref{volume}), one can write
\begin{align}
    \left(\frac{\partial T_H}{\partial P}\right)_{r_h} &= \frac{2r_h}{\sqrt{1+\ell}},\label{dTdP}\\
    \left(\frac{\partial \mathcal{M}}{\partial P}\right)_{r_h} &= V = \frac{4\pi r_h^3}{3\sqrt{1+\ell}}.\label{dMdP}
\end{align}
Applying the chain rule,
\begin{equation}
    \mu_{JT} = \frac{(\partial T_H/\partial P)_{r_h}}{(\partial T_H/\partial r_h)_P\cdot(\partial r_h/\partial P)_{\mathcal{M}}+(\partial T_H/\partial P)_{r_h}}.
\end{equation}
More directly, from the implicit function theorem applied to $\mathcal{M}(r_h,P)=\text{const.}$:
\begin{equation}
\left(\frac{\partial r_h}{\partial P}\right)_{\mathcal{M}} = -\frac{(\partial\mathcal{M}/\partial P)_{r_h}}{(\partial\mathcal{M}/\partial r_h)_P}.
\end{equation}
Noting that $(\partial\mathcal{M}/\partial r_h)_P = T_H\cdot 2\pi r_h \cdot (1+\ell)^{-1/2} \cdot \pi^{-1/2}/(2\sqrt{S/\pi})$ reduces to $T_H\cdot C_p^{-1}\cdot V/(2r_h)$, one arrives at the compact expression
\begin{equation}
    \mu_{JT} = \frac{1}{C_p}\left(\frac{2r_h T_H}{\sqrt{1+\ell}} - V\right).\label{JT-explicit}
\end{equation}
Substituting $V=4\pi r_h^3/(3\sqrt{1+\ell})$ and $T_H$ from~(\ref{temperature}):
\begin{align}
    \mu_{JT} &= \frac{1}{C_p\sqrt{1+\ell}}\left[2r_h T_H - \frac{4\pi r_h^3}{3}\right]\nonumber\\
    &= \frac{r_h}{2\pi C_p\sqrt{1+\ell}}\left[\frac{1-\alpha-\dfrac{\sigma q^2}{r_h^2}+8\pi P r_h^2}{2} - \frac{8\pi^2 r_h^2}{3}\right].\label{JT-full}
\end{align}
The sign of $\mu_{JT}$ determines whether the expansion is a cooling ($\mu_{JT}>0$) or heating ($\mu_{JT}<0$) process. The inversion curve, defined by $\mu_{JT}=0$, separates these two regimes.

\subsection{Inversion Temperature and Inversion Curves}

Setting $\mu_{JT}=0$ in Eq.~(\ref{JT-explicit}) gives the inversion condition:
\begin{equation}
    2r_h T_i = \sqrt{1+\ell}\,V = \frac{4\pi r_h^3}{3},
\end{equation}
where $T_i$ denotes the inversion temperature. Solving for $T_i$:
\begin{equation}
    T_i = \frac{2\pi r_h^2}{3\sqrt{1+\ell}/\sqrt{1+\ell}} = \frac{2\pi r_h^2}{3}.
\end{equation}
More carefully, substituting the explicit form of $T_H$~(\ref{temperature}) into the condition $2r_h T_i / \sqrt{1+\ell} = 4\pi r_h^3/3$:
\begin{equation}
    \frac{1-\alpha-\dfrac{\sigma q^2}{r_h^2}+8\pi P_i r_h^2}{4\pi r_h\sqrt{1+\ell}}\cdot\frac{2r_h}{\sqrt{1+\ell}} = \frac{4\pi r_h^3}{3\sqrt{1+\ell}},
\end{equation}
which simplifies to
\begin{equation}
    1-\alpha - \frac{\sigma q^2}{r_h^2} + 8\pi P_i r_h^2 = \frac{8\pi r_h^2(1+\ell)}{3}.
\end{equation}
Solving for the inversion pressure $P_i$:
\begin{equation}
    P_i = \frac{1}{8\pi r_h^2}\left[\frac{8\pi(1+\ell) r_h^2}{3} - (1-\alpha) + \frac{\sigma q^2}{r_h^2}\right].\label{Pi}
\end{equation}
The inversion temperature is obtained by substituting~(\ref{Pi}) back into~(\ref{temperature}):
\begin{align}
    T_i &= \frac{1}{4\pi r_h\sqrt{1+\ell}}\left[\frac{8\pi(1+\ell) r_h^2}{3} + \frac{\sigma q^2}{r_h^2}\right.\nonumber\\
    &\quad\left.+ (1-\alpha) - \frac{2\sigma q^2}{r_h^2} - (1-\alpha)\right]\nonumber\\
    &= \frac{1}{4\pi r_h\sqrt{1+\ell}}\left[\frac{8\pi(1+\ell) r_h^2}{3} - \frac{\sigma q^2}{r_h^2}\right]\nonumber\\
    &= \frac{2(1+\ell)r_h}{3\sqrt{1+\ell}} - \frac{\sigma q^2}{4\pi r_h^3\sqrt{1+\ell}},
\end{align}
which gives
\begin{equation}
    T_i = \frac{\sqrt{1+\ell}}{4\pi r_h}\left[\frac{8\pi r_h^2}{3} - \frac{\sigma q^2}{r_h^2(1+\ell)}\right].\label{Ti}
\end{equation}
The pair $(T_i, P_i)$ parametrized by $r_h$ traces the inversion curve in the $T$--$P$ plane. As $r_h\to\infty$ (large black holes, high pressure), the charge term in~(\ref{Ti}) becomes negligible and the inversion temperature grows linearly with $r_h$, while as $r_h\to 0$ the charge term dominates and $T_i\to -\infty$, indicating that below a minimum inversion temperature $T_i^{\rm min}$ the black hole always heats upon expansion.

The minimum inversion temperature is found by setting $dT_i/dr_h=0$:
\begin{align}
    \frac{dT_i}{dr_h}&=\frac{\sqrt{1+\ell}}{4\pi}\left[\frac{16\pi r_h}{3}+\frac{4\sigma q^2}{r_h^3(1+\ell)}\right]\frac{1}{r_h}\nonumber\\
    &\quad -\frac{\sqrt{1+\ell}}{4\pi r_h^2}\left[\frac{8\pi r_h^2}{3}-\frac{\sigma q^2}{r_h^2(1+\ell)}\right]=0.
\end{align}
Solving this equation gives the horizon radius $r_h^{\rm min}$ at which the inversion temperature is minimized, and subsequently $T_i^{\rm min}=T_i(r_h^{\rm min})$.

\textbf{Ratio of minimum inversion to critical temperature.} A classical result for the Van der Waals gas is $T_i^{\rm min}/T_c=1/2$. In RN-AdS black holes this ratio is also $T_i^{\rm min}/T_c=1/2$ \cite{Okcu2017}. For our black hole, the ratio $T_i^{\rm min}/T_c$ will depend on $\ell$ and $\alpha$, and its deviation from $1/2$ provides a further quantitative measure of the effects of the cloud of strings and Lorentz violation.

\begin{figure*}[t]
\centering
\includegraphics[width=0.98\textwidth]{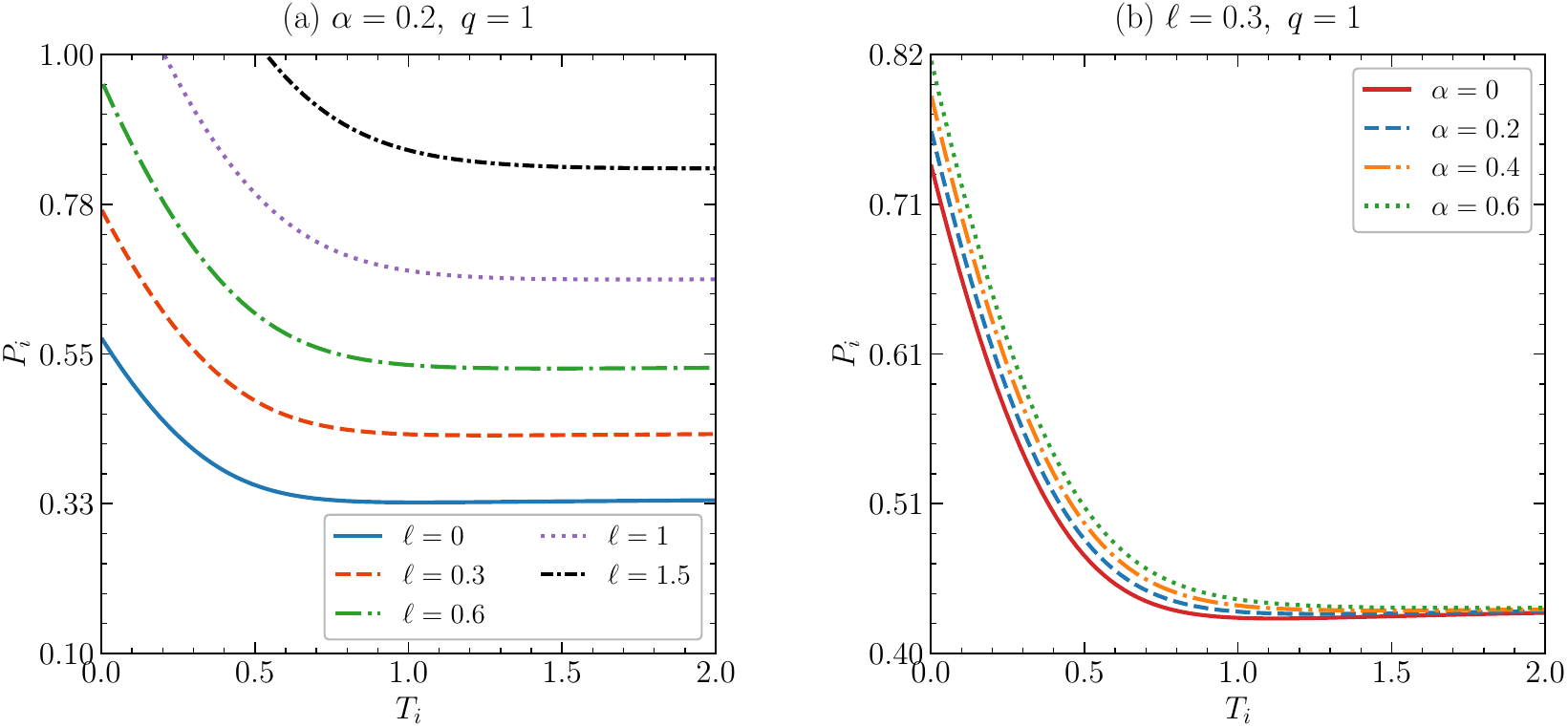}
\caption{Joule--Thomson inversion curves in the $P_i$--$T_i$ plane. Panel (a) shows the effect of the Lorentz-violating parameter $\ell$ for fixed $\alpha=0.2$ and $q=1$, while panel (b) shows the effect of the string-cloud parameter $\alpha$ for fixed $\ell=0.3$ and $q=1$. The curves are plotted from the parametric expressions in Eqs.~(\ref{Pi}) and (\ref{Ti}). They delimit the regions where the sign of the Joule--Thomson coefficient changes within the thermodynamic description adopted here.}
\label{fig:JTinv}
\end{figure*}

The inversion curves obtained from Eqs.~(\ref{Pi}) and (\ref{Ti}) are displayed in Fig.~\ref{fig:JTinv}. They explicitly show that the locus separating heating and cooling regimes depends on $\ell$ and $\alpha$ in the present parametrization.

Figure~\ref{fig:JTinv}(a) indicates that the Lorentz-violating parameter shifts the inversion line appreciably, while Fig.~\ref{fig:JTinv}(b) shows that the string cloud also modifies the inversion locus, though in a quantitatively different way. In both cases, the deformation parameters scale the inversion curve, reflecting that the onset of heating or cooling is sensitive to both the background symmetry-breaking sector and the matter distribution surrounding the black hole.

\subsection{Isenthalpic Curves}

Isenthalpic (constant-mass) curves in the $T_H$--$P$ plane can be obtained by expressing $T_H$ as a function of $P$ at fixed $\mathcal{M}$. From Eq.~(\ref{mass3}), fixing $\mathcal{M}$ implicitly relates $r_h$ to $P$. Differentiating $\mathcal{M}(r_h,P)=\text{const.}$ gives
\begin{equation}
    \left(\frac{\partial r_h}{\partial P}\right)_\mathcal{M} = -\frac{4\pi r_h^3/3}{(1-\alpha)/2 - \sigma q^2/(2r_h^2) + 4\pi P r_h^2}.\label{isenthalpic}
\end{equation}
The slope of an isenthalpic curve at a point $(T_H, P)$ is $\mu_{JT}$ given in~(\ref{JT-full}). The inversion curve passes through the locus of maxima of these isenthalpic curves: to the left (low $P$) of the inversion curve, the slope $\mu_{JT}>0$ (cooling region); to the right (high $P$), $\mu_{JT}<0$ (heating region). The physical picture is that at sufficiently high pressure, the black hole is in a ``compressed'' state where expansion leads to heating, while at low pressure expansion leads to cooling, analogous to the behavior above and below the inversion temperature of a real gas.

\begin{figure*}[t]
\centering
\includegraphics[width=0.98\textwidth]{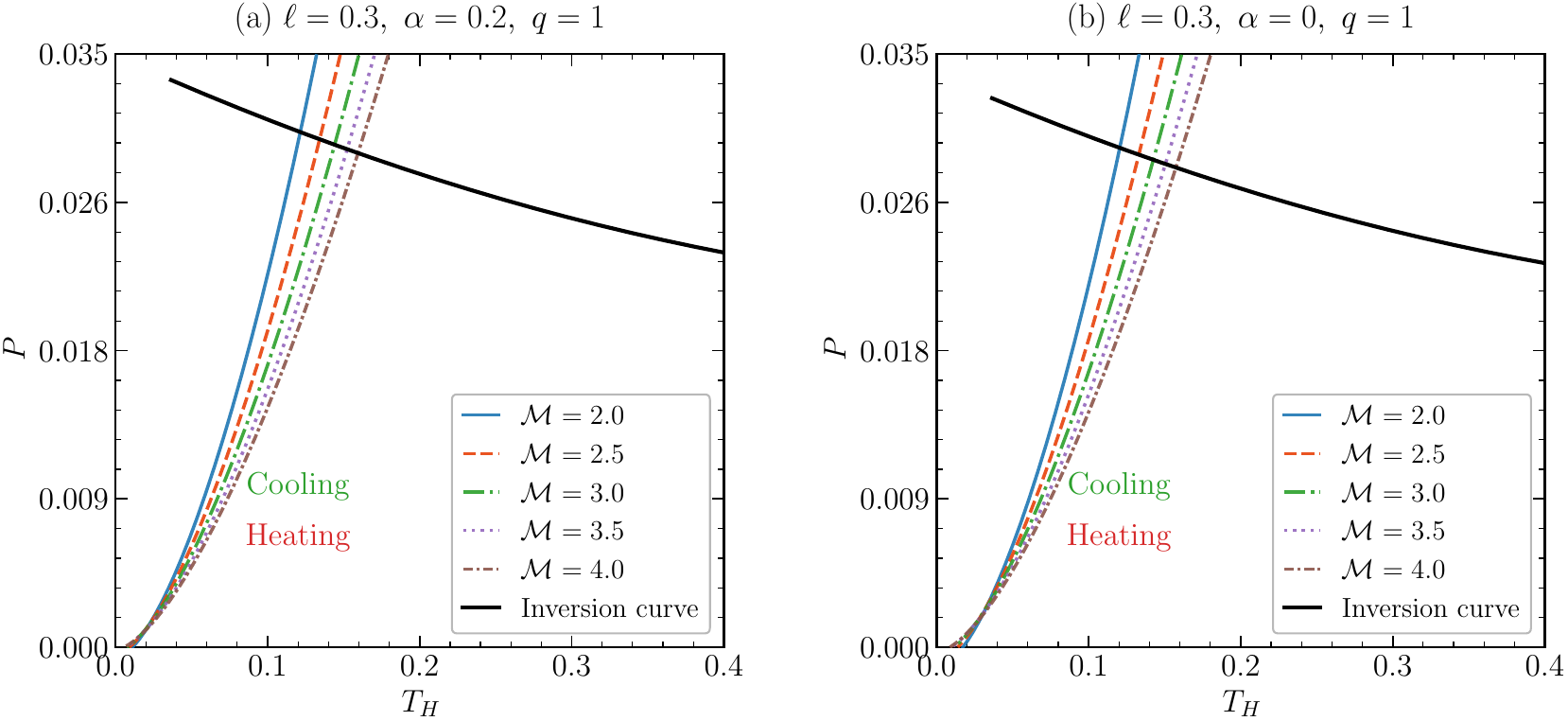}
\caption{Isenthalpic curves in the $P$--$T_H$ plane together with inversion curves. In panel (a), several constant-mass curves are shown for fixed $\ell=0.3$, $\alpha=0.2$, and $q=1$, together with additional inversion curves for different values of $\ell$. In panel (b), the isenthalpic curves are shown for fixed $\ell=0.3$, $\alpha=0$, and $q=1$, together with inversion curves for different values of $\alpha$. The maxima of the isenthalpic curves lie on the inversion line, separating the heating and cooling regions.}
\label{fig:isenthalpic}
\end{figure*}

This behavior is summarized in Fig.~\ref{fig:isenthalpic}, where representative isenthalpic curves are superposed on the inversion line. The maxima of the isenthalpic trajectories occur precisely on the inversion boundary that separates the heating and cooling sectors.

Figure~\ref{fig:isenthalpic} makes the thermodynamic interpretation especially transparent. On one side of the inversion line, the isenthalpic slope is positive, and the black hole cools as the pressure decreases; on the other side, the slope is negative, and the system heats up during expansion. The displacement of the inversion line caused by $\ell$ and $\alpha$, therefore, has direct consequences for the domain in which each thermodynamic response occurs.

\section{Thermodynamics from Tsallis Entropy-based Framework}\label{sec:5}

Considering black holes as gravitational thermodynamic systems with long-range interactions, the standard Bekenstein--Hawking entropy $S_{\rm BH}=A/4$ may not fully capture the non-extensive nature of the entropy of the system. This issue was addressed in Ref.~\cite{Tsallis2013}, where a non-additive (Tsallis-type) entropic functional was proposed for systems exhibiting long-range correlations, such as gravitational systems and black holes.

\subsection{Tsallis Entropy and Modified Mass}

The Tsallis entropy for a black hole is defined as~\cite{Tsallis2013}
\begin{equation}
S_{\delta}\equiv\tilde{S}=\left(\frac{A_{\rm BH}}{4}\right)^{\delta}=(\pi r_h^2)^{\delta},
\label{tsallis}
\end{equation}
where $\delta\in(0,3/2]$ is the non-extensivity (Tsallis) parameter governing the deviation from standard thermodynamics. In the limit $\delta\to 1$, $\tilde{S}\to S_{\rm BH}=\pi r_h^2$, recovering the usual Bekenstein--Hawking result. For $\delta\neq 1$, the entropy is a nonlinear power of the horizon area, encoding corrections arising from the non-extensive nature of gravitational entropy.

Since $\tilde{S}=(\pi r_h^2)^\delta$, we have $r_h^2=\tilde{S}^{1/\delta}/\pi$, i.e., $S\to\tilde{S}^{1/\delta}$ in all formulas derived in Sec.~\ref{sec:3}. The ADM mass~(\ref{mass4}) in terms of the Tsallis entropy becomes
\begin{equation}
\mathcal{M}=\frac{\left[(1-\alpha)\,\tilde{S}^{\frac{1}{2\delta}}+\sigma \pi q^2\,\tilde{S}^{-\frac{1}{2\delta}}+\frac{8P}{3}\,\tilde{S}^{\frac{3}{2\delta}}\right]}{2\sqrt{(1+\ell) \pi}}.\label{mass5}
\end{equation}

\subsection{Temperature and Specific Heat in Tsallis Framework}

The black hole temperature in the Tsallis framework is obtained from the first law as $T(\tilde{S})=(\partial\mathcal{M}/\partial\tilde{S})_P$. 

Carrying out the full derivative of the ADM mass $\mathcal{M}$ and collecting terms, one finds the black hole temperature as,
\begin{align}
    T(\tilde{S})&=\frac{1}{4\delta\sqrt{(1+\ell) \pi}}\Bigl[(1-\alpha)\,\tilde{S}^{\frac{1}{2\delta}-1}\nonumber\\
    &\quad - \sigma\pi q^2\,\tilde{S}^{-\frac{1}{2\delta}-1} + 8P\,\tilde{S}^{\frac{3}{2\delta}-1}\Bigr].\label{temperature3}
\end{align}

In the limit $\delta\to 1$, one recovers the standard Hawking temperature~(\ref{temperature2}) upon re-inserting $\tilde{S}=S=\pi r_h^2$. For $\delta\neq 1$, both the temperature and its dependence on $\tilde{S}$ are modified: in particular, the effective power-law scaling of $T$ with the horizon area changes, which can significantly alter the thermodynamic behavior at small horizon sizes.

The specific heat at constant pressure in the Tsallis framework is
\begin{equation}
C_p(\tilde{S})=T(\tilde{S})\left(\frac{\partial\tilde{S}}{\partial T}\right)_P =T(\tilde{S})\left(\frac{\partial T}{\partial \tilde{S}}\right)^{-1}_P,
\end{equation}
which, after substituting Eqs.~(\ref{mass5}) and~(\ref{temperature3}), yields
\begin{widetext}
\begin{align}
C_p(\tilde{S})=\frac{(1-\alpha)\,\tilde{S}^{\frac{1}{2\delta}-1} - \sigma\pi q^2\,\tilde{S}^{-\frac{1}{2\delta}-1} + 8P\,\tilde{S}^{\frac{3}{2\delta}-1}}
{(1-\alpha)\!\left(\frac{1}{2\delta}-1\right)\!\tilde{S}^{\frac{1}{2\delta}-2} + \sigma\pi q^2\!\left(\frac{1}{2\delta}+1\right)\!\tilde{S}^{-\frac{1}{2\delta}-2} + 8P\!\left(\frac{3}{2\delta}-1\right)\!\tilde{S}^{\frac{3}{2\delta}-2}}.\label{heat-3}
\end{align}
\end{widetext}
The numerator of~(\ref{heat-3}) is proportional to $T(\tilde{S})$, so that $C_p$ changes sign only when the denominator vanishes, defining the critical (Davies) points of the Tsallis-modified black hole. For $\delta=1$, Eq.~(\ref{heat-3}) reduces to the earlier result~(\ref{heat-1}) when expressed in terms of $r_h$. For $\delta\neq 1$, the additional power-law factors in the denominator shift the positions of the Davies points and can introduce new stability windows not present in the standard ($\delta=1$) case.

\begin{figure*}[t]
\centering
\includegraphics[width=0.98\textwidth]{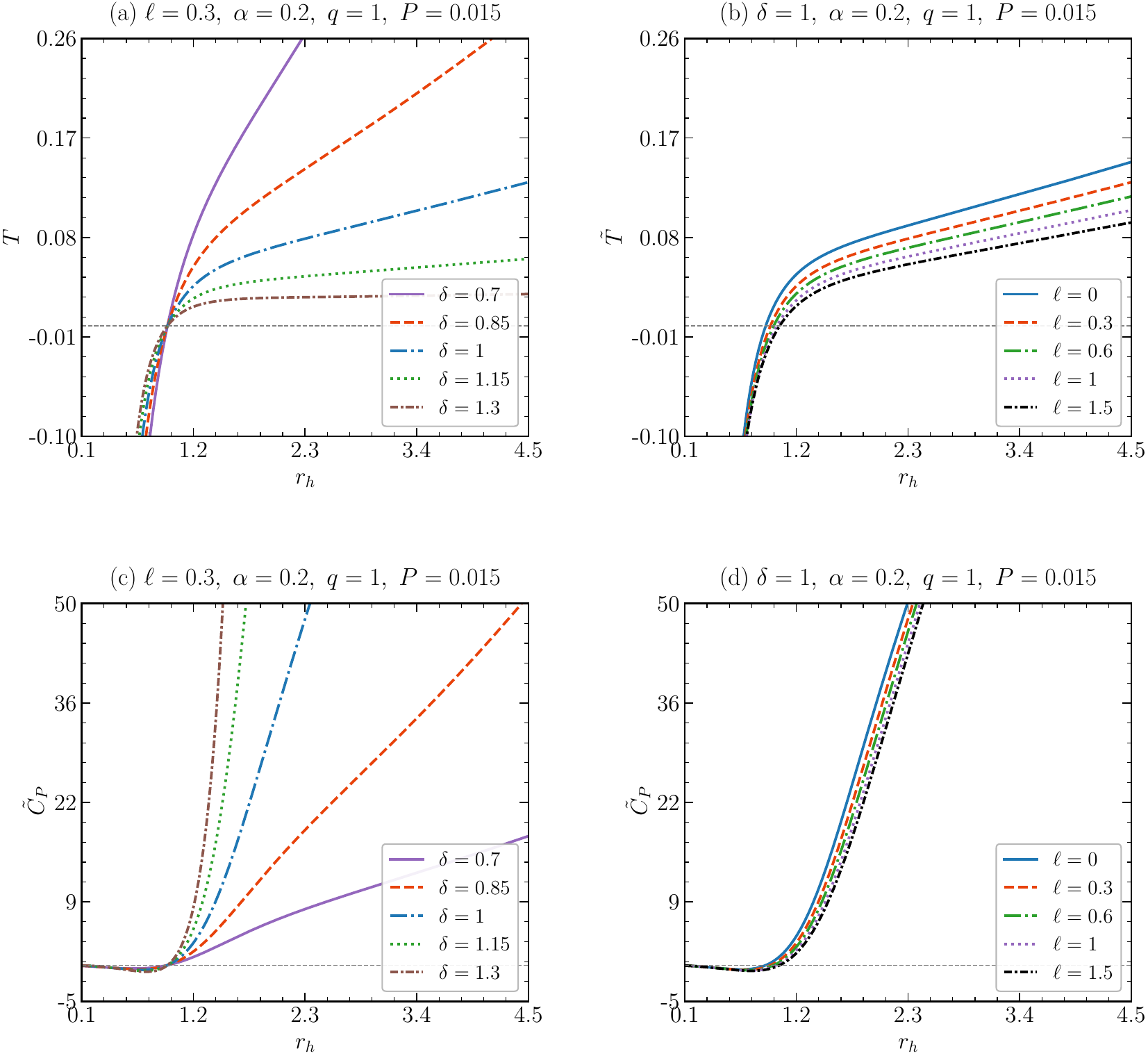}
\caption{Thermodynamic quantities in the Tsallis entropy-based framework. Panel (a) shows the modified temperature $\tilde{T}$ as a function of $r_h$ for different values of the non-extensivity parameter $\delta$ at fixed $\ell=0.3$, $\alpha=0.2$, $q=1$, and $P=0.015$. Panel (b) shows $\tilde{T}$ as a function of $r_h$ for different values of $\ell$ at fixed $\delta=1$, $\alpha=0.2$, $q=1$, and $P=0.015$. Panels (c) and (d) show the corresponding modified specific heat $\tilde{C}_P$ for the same parameter choices.}
\label{fig:tsallis}
\end{figure*}

The effect of non-extensivity on the temperature and specific heat is shown in Fig.~\ref{fig:tsallis}. This figure allows one to compare the standard case $\delta=1$ with sub-extensive and super-extensive regimes.

Figure~\ref{fig:tsallis}(a) shows that increasing $\delta$ changes the shape and scale of the temperature profile, especially in the small-horizon regime where non-extensive effects are strongest. The corresponding specific-heat curves in Fig.~\ref{fig:tsallis}(c) reveal that the stability windows and divergence points are also displaced as $\delta$ changes. Thus, Tsallis entropy modifies not only the thermal scale but also the topology of the stability diagram. Panels (b) and (d) show that, once the Tsallis framework is fixed, the Lorentz-violating parameter $\ell$ still plays a major role by shifting both the temperature and the Davies points.

\subsection{Equation of State and Criticality in Tsallis Framework}

Substituting $\tilde{S}=\left(\pi r_h^2\right)^\delta$ and using the identification $v=2r_h$, Eq.~(\ref{temperature3}) can be rearranged to yield the modified equation of state:
\begin{equation}
    P=c_0\,\frac{T}{v^{3-2\delta}}-\frac{c_1}{v^2}+\frac{c_2}{v^4},\label{critical-8}
\end{equation}
where
\begin{equation}
    c_0=\frac{\delta\sqrt{1+\ell}\,\pi^{\delta-1}}{2^{2(\delta-1)}},\label{critical-9}
\end{equation}
and $c_1$, $c_2$ are as in~(\ref{critical-4}). 

For $\delta \to 1$, $c_0\to\sqrt{1+\ell}$ and~(\ref{critical-8}) reduces to~(\ref{critical-3}), as expected. The non-trivial new feature for $\delta\neq 1$ is the $v$-dependent prefactor $v^{3-2\delta}$ multiplying the thermal term. For $\delta>1$ (super-extensive entropy), the thermal pressure falls off more slowly with $v$ than in the standard case, while for $\delta<1$ it falls off faster, effectively modifying the range of the repulsive thermal interaction.

The critical point in the Tsallis framework is again determined by~(\ref{critical-5}) applied to~(\ref{critical-8}):
\begin{align}
\frac{\partial P}{\partial v}\bigg|_T &= -(3-2\delta)\,c_0\,\frac{T}{v^{4-2\delta}}+\frac{2c_1}{v^3}-\frac{4c_2}{v^5}=0,\label{dPdv-T}\\
\frac{\partial^2 P}{\partial v^2}\bigg|_T &= (3-2\delta)(4-2\delta)\,c_0\,\frac{T}{v^{5-2\delta}}-\frac{6c_1}{v^4}+\frac{20c_2}{v^6}=0.\label{d2Pdv2-T}
\end{align}
Eliminating $T$ from~(\ref{dPdv-T}) and~(\ref{d2Pdv2-T}) by multiplying~(\ref{dPdv-T}) by $(4-2\delta)/v$ and subtracting from~(\ref{d2Pdv2-T}), one obtains
\begin{equation}
    \frac{(2+4\delta-4\delta^2)c_1}{v_c^4} - \frac{(4+8\delta)c_2}{v_c^6} = 0,
\end{equation}
which gives
\begin{align}
v_c &= \sqrt{\frac{2(1+2\delta)\,c_2}{(2\delta-1)\,c_1}}\nonumber\\
&=\sqrt{\frac{2(1+2\delta)}{(2\delta-1)}\,\frac{4q^2(1+\ell)}{(1-\alpha)(1+\ell/2)}}.\label{critical-10}
\end{align}
The critical temperature and pressure are then
\begin{align}
T_c &= \frac{2c_1 v_c^{1-2\delta} - 4c_2 v_c^{-1-2\delta}}{c_0(3-2\delta)},\label{critical-11}\\[2mm]
P_c &= c_0\,\frac{T_c}{v_c^{3-2\delta}}-\frac{c_1}{v_c^2}+\frac{c_2}{v_c^4}.\label{critical-12}
\end{align}
One can verify that Eqs.~(\ref{critical-10})--(\ref{critical-12}) reduce to~(\ref{vc})--(\ref{Pc}) in the limit $\delta\to 1$. Note that for $\delta=1/2$, the denominator in~(\ref{critical-10}) vanishes, indicating the absence of a standard Van der Waals-like critical point; this limiting case requires separate treatment and may signal a qualitative change in the phase structure. Moreover, the Tsallis parameter $\delta>1/2$ in order for the specific volume $v_c$ to be real and positive.

The dependence of the critical specific volume~(\ref{critical-10}) on $\delta$ is particularly instructive: increasing $\delta$ (more super-extensive entropy) increases $v_c$, meaning the critical point moves to larger horizon sizes. This is consistent with the physical expectation that non-extensive entropy corrections, which become more important for larger systems, shift the onset of phase transitions to larger scales.

\begin{figure*}[t]
\centering
\includegraphics[width=0.98\textwidth]{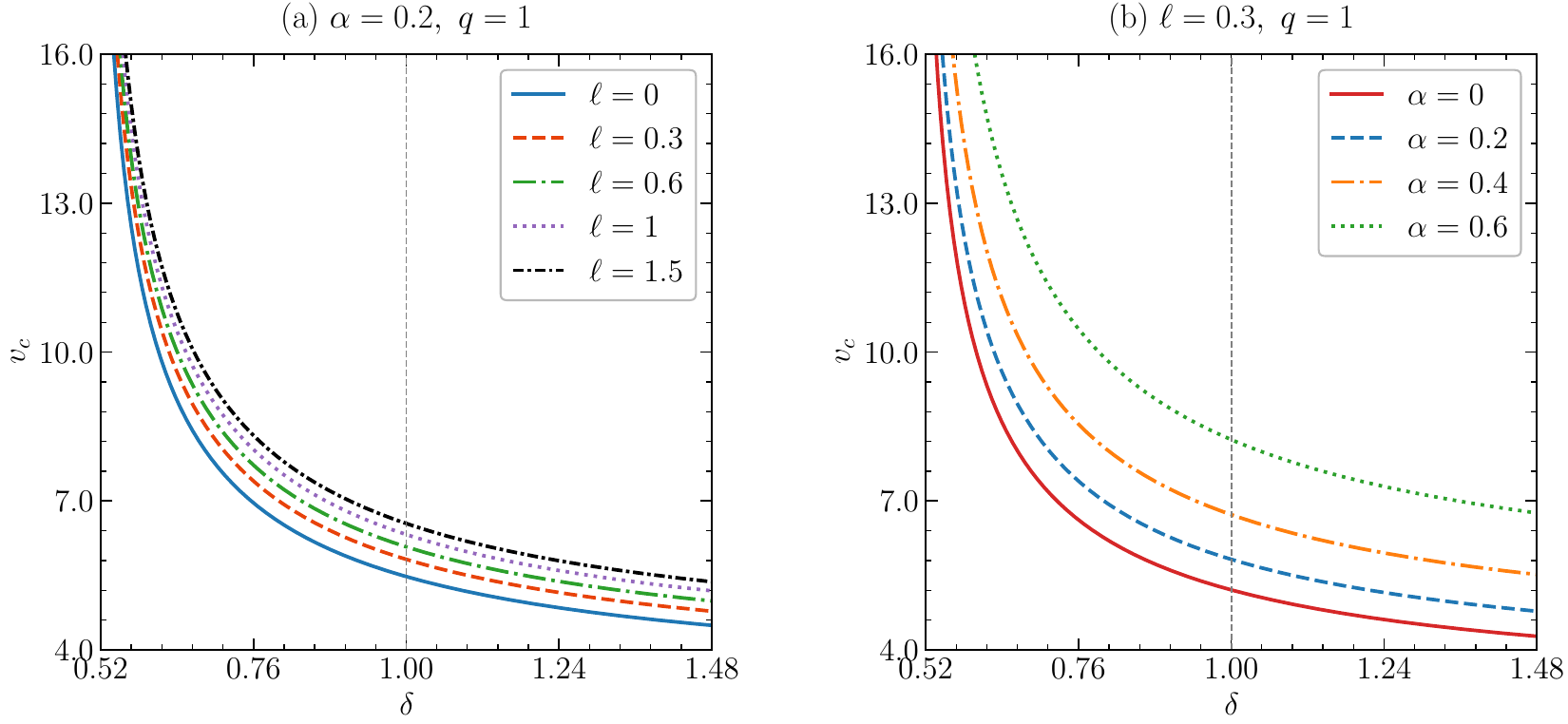}
\caption{Critical specific volume $v_c$ in the Tsallis framework as a function of the non-extensivity parameter $\delta$. Panel (a) shows the dependence on $\delta$ for several values of the Lorentz-violating parameter $\ell$ at fixed $\alpha=0.2$ and $q=1$. Panel (b) shows the corresponding dependence for several values of the cloud-of-strings parameter $\alpha$ at fixed $\ell=0.3$ and $q=1$. The divergence near $\delta=1/2$ reflects the denominator $(2\delta-1)$ in Eq.~(\ref{critical-10}), signaling the breakdown of the standard critical-point structure in that limit.}
\label{fig:tsallisvc}
\end{figure*}

This behavior is shown in Fig.~\ref{fig:tsallisvc}, where the critical volume is plotted as a function of $\delta$ for different values of $\ell$ and $\alpha$.

Figure~\ref{fig:tsallisvc} shows that the critical volume grows monotonically with $\delta$ in the physically relevant domain $\delta>1/2$. This confirms that stronger non-extensive effects move the critical point to larger scales. The figure also shows that both Lorentz violation and the cloud of strings quantitatively amplify or suppress this growth, thereby coupling the Tsallis deformation to the underlying geometry and matter content.

\subsection{Gibbs free energy}

In this part, we determine the Gibbs free energy of the thermodynamic system within the Tsallis entropy-based framework and analyze how the Tsallis parameter modifies it.

The Gibbs free energy in the Tsallis entropy framework can be obtained using the following relation:
\begin{equation}
    G(\tilde{S})=\mathcal{M}(\tilde{S})-\tilde{S}\,T(\tilde{S}).\label{gibbs-1}
\end{equation}

Substituting ADM mass given in (\ref{mass5}) and the black hole temperature (\ref{temperature3}), we obtain
\begin{align}
G(\tilde{S})&=\frac{1}{4\delta\sqrt{(1+\ell)\pi}}
\Big[
(2\delta-1)(1-\alpha)\,\tilde S^{\frac{1}{2\delta}}\nonumber\\
&+ (2\delta+1) \sigma \pi q^2 \,\tilde S^{-\frac{1}{2\delta}}
- \frac{8(3\delta-2)}{3} P\, \tilde S^{\frac{3}{2\delta}}
\Big].
.\label{gibbs-2}
\end{align}
In the limit $\delta \to 1$, we have $\tilde{S} \to S$, and hence, the Gibbs free energy in (\ref{gibbs-2}) simplifies to the result reported in (\ref{gibbs}).

\begin{figure}[t]
\centering
\includegraphics[width=0.98\columnwidth]{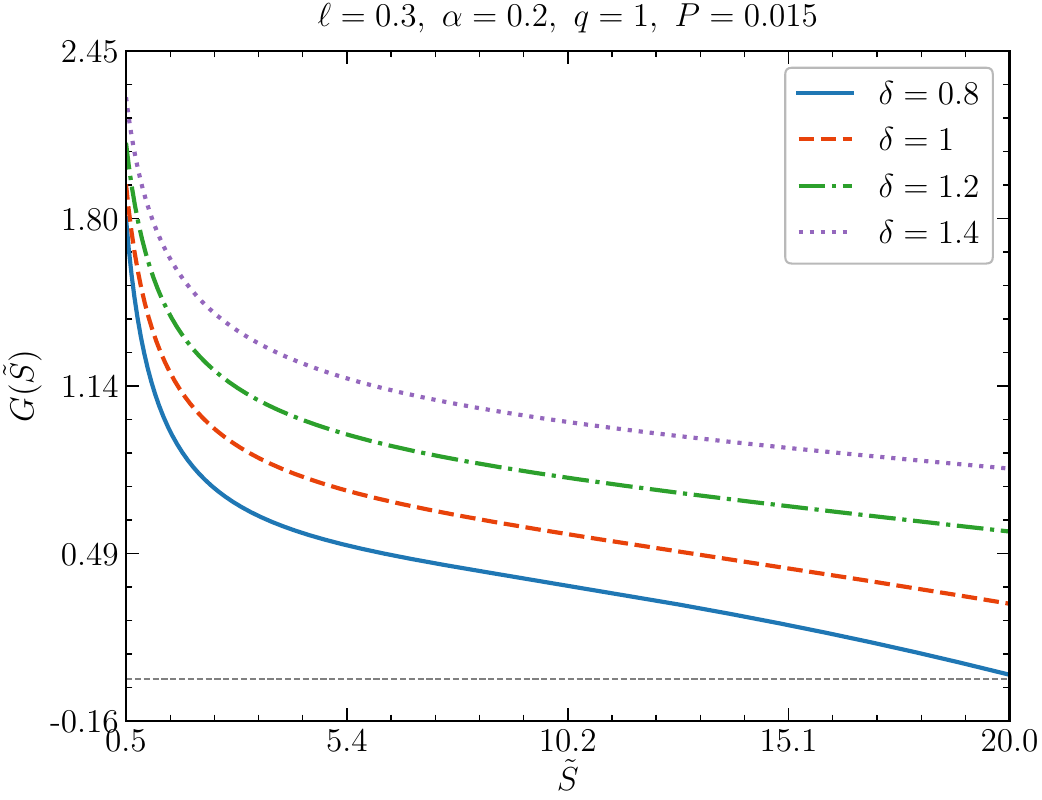}
\caption{Gibbs free energy $G(\tilde S)$ in the Tsallis entropy-based framework as a function of the Tsallis entropy $\tilde S$ for $\delta=0.8,\,1.0,\,1.2,$ and $1.4$, with fixed $\ell=0.3$, $\alpha=0.2$, $q=1$, and $P=0.015$. The curves are plotted from Eq.~(76). The dependence on $\delta$ shows that the non-extensive parameter changes both the overall magnitude of the Gibbs free energy and the entropy scale at which the thermodynamic branches evolve.}
\label{fig:gibbs_tsallis}
\end{figure}

The Gibbs free energy in the Tsallis entropy-based framework is displayed in Fig.~\ref{fig:gibbs_tsallis}. Since $G(\tilde S)$ depends explicitly on the non-extensivity parameter $\delta$, this quantity provides a useful probe of how the global thermodynamic preference of the black-hole branches is modified when the standard extensive entropy is replaced by its Tsallis counterpart.

Figure~\ref{fig:gibbs_tsallis} shows that the Gibbs free energy is strongly sensitive to the Tsallis deformation. As $\delta$ changes, the profiles of $G(\tilde S)$ are displaced and reshaped, indicating that the non-extensive entropy modifies the global thermodynamic structure of the system. Since the Gibbs free energy determines which branch is thermodynamically preferred in a canonical description, these deformations imply that the Tsallis parameter affects the competition among black-hole phases in a nontrivial manner.

Another relevant feature is that the curves remain clearly separated over the whole entropy interval, showing that the effect of $\delta$ is not restricted to a narrow region. In particular, the shift in the sign and slope of $G(\tilde S)$ may alter the location of phase transitions and the range of entropy values for which a given branch becomes globally favored. Therefore, Fig.~\ref{fig:gibbs_tsallis} provides direct evidence that the Tsallis framework changes not only local stability properties but also the global free-energy landscape of the black hole.

From a physical point of view, this means that non-extensive entropy corrections influence the balance between the thermal and interaction contributions encoded in the Gibbs potential. Consequently, the coexistence structure familiar in the standard AdS case may be quantitatively modified upon introducing the Tsallis parameter.

\subsection{Helmholtz Free Energy}

Finally, we determine the Helmholtz free energy of the thermodynamic system within a Tsallis entropy-based framework and analyze how the Tsallis parameter modifies it. This energy can be obtained as
\begin{equation}
    F(\tilde{S})=\mathcal{M}(\tilde{S})-P\,V(\tilde{S}).\label{free-1}
\end{equation}

Now, using the ADM mass in (\ref{mass5}), we find the volume as
\begin{equation}
    V(\tilde{S})=\left(\frac{\partial \mathcal{M}}{\partial P}\right)_{\tilde{S},\alpha,q}
\end{equation}
In our case at hand, we find
\begin{equation}
    V(\tilde{S})=\frac{4}{3\sqrt{(1+\ell)\pi}}\,\tilde{S}^{\frac{3}{2\delta}}.\label{free-2}
\end{equation}

Substituting $\mathcal{M}(\tilde{S})$ and $V(\tilde{S})$ into the relation (\ref{free-1}) results
\begin{align}
F(\tilde{S})=\frac{1}{2\sqrt{(1+\ell)\pi}}
\left[
(1-\alpha)\,\tilde S^{\frac{1}{2\delta}}
+ \sigma \pi q^2 \,\tilde S^{-\frac{1}{2\delta}}
\right].\label{free-3}
\end{align}
In the limit $\delta \to 1$, the Helmholtz free energy in (\ref{free-3}) simplifies to the result obtained in (\ref{free}).

\begin{figure}[t]
\centering
\includegraphics[width=0.98\columnwidth]{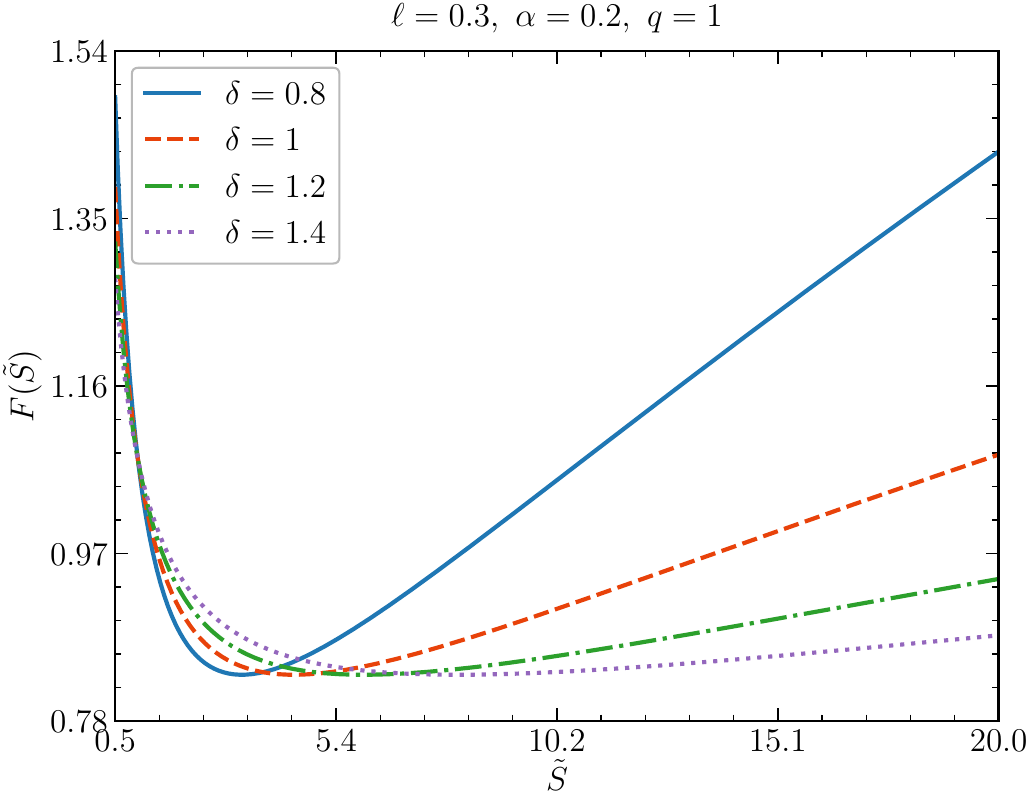}
\caption{Helmholtz free energy $F(\tilde S)$ in the Tsallis entropy-based framework as a function of the Tsallis entropy $\tilde S$ for $\delta=0.8,\,1.0,\,1.2,$ and $1.4$, with fixed $\ell=0.3$, $\alpha=0.2$, and $q=1$. The curves are plotted from Eq.~(80). The separation among the profiles shows that the Tsallis parameter modifies the energetic content of the black hole even in the absence of an explicit pressure contribution in the final expression for $F(\tilde S)$.}
\label{fig:helmholtz_tsallis}
\end{figure}

The Helmholtz free energy in the Tsallis entropy-based framework is shown in Fig.~\ref{fig:helmholtz_tsallis}. In contrast with the Gibbs free energy, the Helmholtz potential is the relevant thermodynamic quantity for fixed-volume descriptions and provides complementary information about the internal energetic cost of the configuration.

As seen in Fig.~\ref{fig:helmholtz_tsallis}, the Helmholtz free energy also depends significantly on the non-extensivity parameter $\delta$. The curves display a systematic separation as $\delta$ varies, indicating that the Tsallis deformation changes the internal thermodynamic balance between the string-cloud contribution and the effective charge sector. This confirms that the non-extensive parameter directly imprints itself on the black hole's free-energy content.

A noteworthy point is that the dependence of $F(\tilde S)$ on $\delta$ persists even though the final expression does not contain the pressure term explicitly. This means that the Tsallis modification is already encoded at a more fundamental level in the entropy scaling itself, thereby affecting the free energy independently of the pressure contribution, which appears more directly in the Gibbs potential. Hence, the effect of $\delta$ on the Helmholtz free energy is conceptually cleaner, since it isolates how the entropy deformation alone changes the system's energetic structure.

Taken together, Figs.~\ref{fig:gibbs_tsallis} and \ref{fig:helmholtz_tsallis} show that the Tsallis parameter modifies both the global thermodynamic preference and the internal energetic content of the black hole. This reinforces the conclusion that non-extensive entropy corrections lead to substantial departures from the standard Bekenstein-Hawking framework.

The thermodynamic analysis of a charged AdS black hole within the Tsallis entropy framework shows that key thermodynamic quantities, such as the black hole temperature, critical points, Gibbs free energy, and Helmholtz free energy, depend not only on the Lorentz-violating parameter $\ell$ and the string cloud parameter $\alpha$ but are also affected by the Tsallis parameter $\delta$. Thus, this parameter $\delta$ modifies the thermodynamic structure relative to the standard results of black hole thermodynamics.

\section{Sparsity}

Sparsity is an important characteristic of black hole (BH) radiation, providing a quantitative measure of how “discrete” the emission process is. It is typically defined as the ratio of the average time interval between the emission of successive quanta to the characteristic timescales set by the energies of the emitted quanta. In other words, sparsity captures how far apart individual Hawking quanta are emitted relative to their typical oscillation or energy timescales.

One of the most striking features of Hawking radiation is that it is extremely sparse throughout the black hole evaporation process, especially when compared to standard blackbody radiation. For a blackbody at a similar temperature, the emission of photons occurs in a nearly continuous stream, whereas Hawking quanta are emitted with long intervals between them. This extreme sparsity has important implications: it allows for a clear distinction between Hawking radiation and classical thermal radiation, and it affects the detectability and temporal correlations of emitted particles. Moreover, sparsity is directly related to the semiclassical nature of Hawking radiation, reflecting the fact that the black hole horizon radiates only a few quanta at a time, rather than a dense flux.

Several studies have explored the consequences of this sparsity, showing that it influences not only the emission spectrum but also the statistical and thermodynamic properties of evaporating black holes \cite{Hawking1975,Page1976,Gray2016}. In particular, sparsity can affect the rate of information release during evaporation and plays a role in theoretical considerations of black hole remnants and the black hole information paradox. Therefore, understanding and quantifying sparsity is essential for a complete description of black hole thermodynamics and quantum radiation processes.

Sparsity can be quantified by introducing the parameter
\begin{equation}
    \eta=\frac{C}{\tilde{g}}\,\frac{\lambda^2_t}{A_{\rm eff}}.\label{sparsity-1}
\end{equation}
Here, $C$ is a dimensionless constant, $\tilde{g}$ denotes the spin degeneracy factor of the emitted quanta, $\lambda_t=2\pi/T$ represents the thermal wavelength associated with the radiation, $T$ is the black hole temperature, and $A_{\rm eff}=27 A_{\rm BH}/4$ corresponds to the effective area of the black hole. For the case of selected BH in the Tsallis entropy framework, we have \(A_{\rm eff}=27\tilde{S}^{1/\delta}\).

\begin{figure}[t]
\centering
\includegraphics[width=0.98\columnwidth]{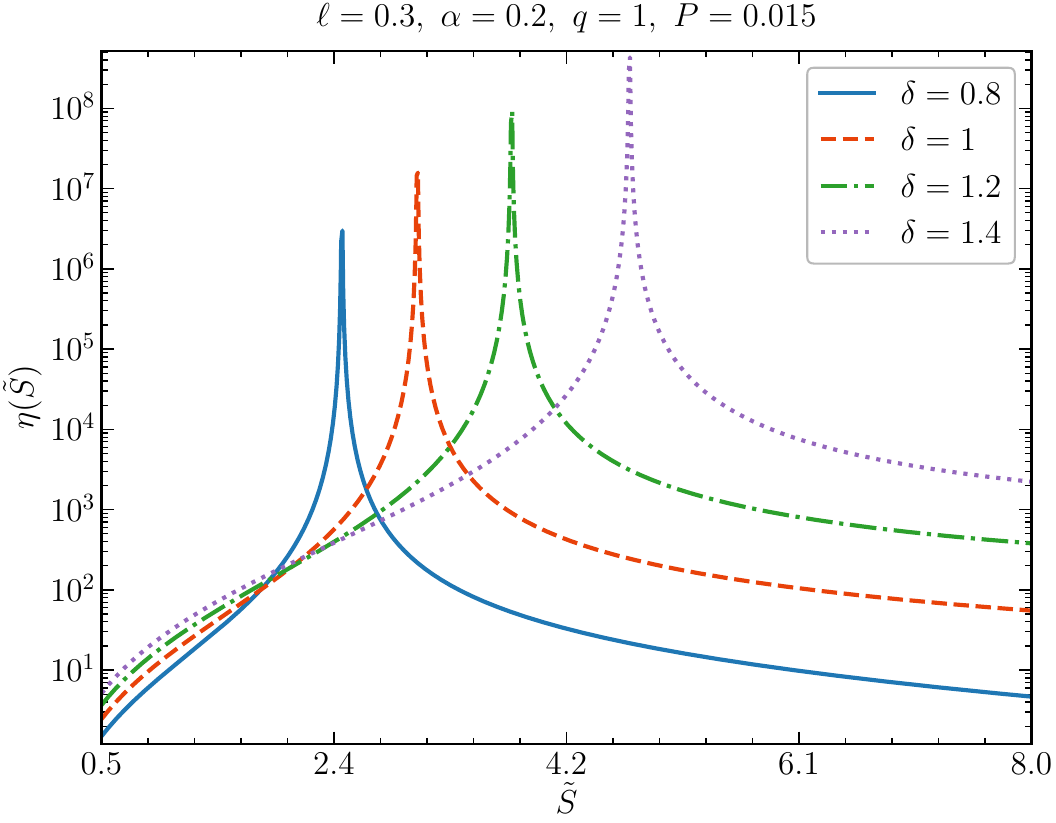}
\caption{Sparsity parameter $\eta(\tilde S)$ as a function of the Tsallis entropy $\tilde S$ for $\delta=0.8,\,1.0,\,1.2,$ and $1.4$, with fixed $\ell=0.3$, $\alpha=0.2$, $q=1$, and $P=0.015$. The vertical axis is shown on a logarithmic scale in order to make visible the large variation of $\eta(\tilde S)$ across the parameter range. As $\delta$ increases, the position and height of the peaks are shifted, indicating that the non-extensive parameter strongly modifies the sparsity of the Hawking emission. The sharp peaks occur when the denominator of Eq.~(\ref{sparsity-2}) becomes very small, leading to a strong enhancement of $\eta(\tilde S)$.}
\label{fig:eta_tsallis}
\end{figure}

Therefore, the sparsity parameter in the Tsallis entropy-based framework is given by ($\eta_{\rm Sch.}=64\pi^3/27$)
\begin{align}
    \eta \left(\tilde{S}\right)=\delta^2 (1+\ell)\Big[&(1-\alpha)\tilde{S}^{\frac{1}{\delta}-1}- \sigma\pi q^2\,\tilde{S}^{-1} \nonumber\\
    &+ 8P\,\tilde{S}^{\frac{2}{\delta}-1}\Big]^{-2}\,\eta_{\rm Sch.}.
    \label{sparsity-2}
\end{align}
The behavior of the sparsity parameter in the Tsallis framework is illustrated in Fig.~\ref{fig:eta_tsallis}. Since the quantity $\eta(\tilde S)$ defined in Eq.~(\ref{sparsity-2}) depends explicitly on the non-extensivity parameter $\delta$, it provides a useful diagnostic of how the Hawking cascade is modified when the standard Bekenstein--Hawking entropy is replaced by its Tsallis counterpart. In particular, the logarithmic scale on the vertical axis allows visualization of both the regular regions and the sharp enhancements associated with the zeros of the denominator in Eq.~(\ref{sparsity-2}).

Figure~\ref{fig:eta_tsallis} shows that the sparsity parameter is highly sensitive to the Tsallis deformation. For all values of $\delta$ considered here, $\eta(\tilde S)$ increases rapidly as $\tilde S$ approaches a critical region, reaches a pronounced maximum, and then decreases again away from that region. This non-monotonic behavior reflects the structure of Eq.~(\ref{sparsity-2}), where $\eta(\tilde S)$ is inversely proportional to the square of an effective thermodynamic combination involving $(1-\alpha)$, the charge contribution, and the pressure term. Whenever this combination becomes small, the corresponding Hawking cascade becomes extremely sparse, which explains the large peaks observed in the figure.

Another important feature is that the location of the peak moves systematically as $\delta$ changes. For smaller values of $\delta$, the enhancement occurs at lower values of $\tilde S$, whereas for larger $\delta$ the peak is displaced toward larger $\tilde S$. At the same time, the peak magnitude also changes substantially, indicating that the non-extensive parameter not only shifts the characteristic entropy scale of the emission process but also modifies its intensity. In this sense, $\delta$ acts as a direct control parameter for the sparsity of the Hawking radiation in the Tsallis-based description.

The logarithmic representation is especially useful here because it reveals that the differences among the curves are not restricted to the immediate vicinity of the peaks. Even away from the critical regions, the four profiles remain clearly separated, demonstrating that the Tsallis parameter leaves a measurable imprint on the global behavior of $\eta(\tilde S)$. Therefore, Fig.~\ref{fig:eta_tsallis} provides direct evidence that the non-extensive entropy framework leads to substantial modifications in the radiation sparsity when compared with the standard $\delta=1$ case.

From a physical perspective, these results indicate that the interplay among Lorentz violation, the cloud-of-strings sector, and the Tsallis entropy correction produces a richer evaporation pattern than in the usual extensive framework. In particular, the strong sensitivity of $\eta(\tilde S)$ to $\delta$ suggests that sparsity observables may serve as an additional probe of non-extensive effects in black-hole thermodynamics.

In terms of the horizon radius $r_h$, the expression (\ref{sparsity-2}) can be written as
\begin{align}
\eta \left(r_h\right)&=\delta^2 (1+\ell)\Big[(1-\alpha) \pi^{1-\delta} r^{2-2\delta}_h\nonumber\\
&- \sigma \pi^{1-\delta} q^2 r^{-2\delta}_h
+ 8 P \pi^{2-\delta} r^{2(2-\delta)}_h\Big]^{-2}\eta_{\rm Sch.}.
\label{sparsity-3}
\end{align}

In the limit $\delta \to 1$, the sparsity parameter in the Bekenstein-Hawking entropy framework for the considered AdS black hole reduces to
\begin{align}
    \eta \left(r_h\right)=(1+\ell)\Big[1-\alpha- \sigma q^2/r^{2}_h+ 8 P \pi r^{2}_h\Big]^{-2}\eta_{\rm Sch.}.
    \label{sparsity-4}
\end{align}

\begin{figure}[t]
\centering
\includegraphics[width=0.98\columnwidth]{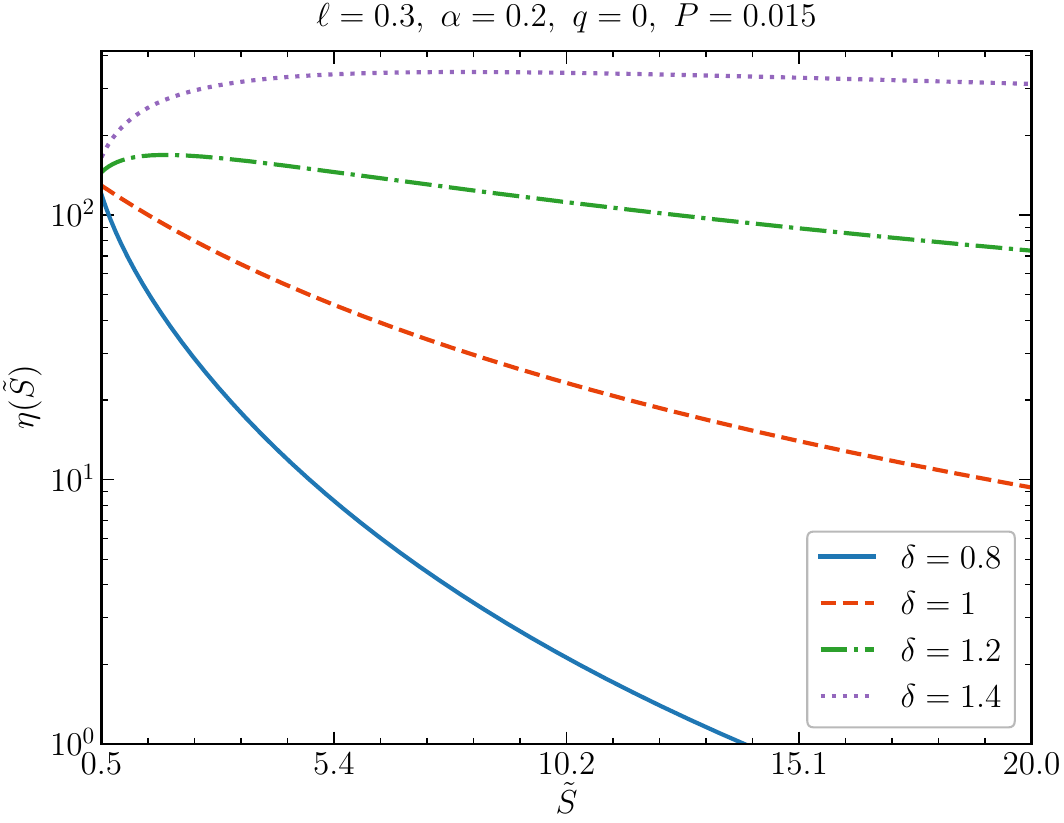}
\caption{Sparsity parameter $\eta(\tilde S)$ as a function of the Tsallis entropy $\tilde S$ for the uncharged black hole case ($q=0$), with $\delta=0.8,\,1.0,\,1.2,$ and $1.4$, and fixed $\ell=0.3$, $\alpha=0.2$, and $P=0.015$. The vertical axis is shown on a logarithmic scale in order to make visible the broad variation of $\eta(\tilde S)$ over the chosen entropy range. In the absence of the charge term, the profiles become smoother, and no sharp divergence associated with the electric contribution is present. The curves remain clearly separated, showing that the Tsallis parameter $\delta$ strongly affects the sparsity of the Hawking emission even in the neutral sector.}
\label{fig:eta_tsallis_uncharged}
\end{figure}

Furthermore, in the limit $q=0$ and $P=0$, the considered spacetime reduces to the Schwarzschild black hole surrounded by a cloud of strings (known as Letelier solution) in bumblebee gravity \cite{Casana2018}. In this case, the sparsity parameter obtained from Eq.~(\ref{sparsity-3}) becomes
\begin{align}
\eta=\delta^2 (1+\ell)( 1-\alpha)^{2(1-2 \delta)}\pi^{2(\delta-1)}\left(2M\right)^{4 (\delta-1)}\eta_{\rm Sch.},
    \label{sparsity-5}
\end{align}
where we have used the horizon radius $r_h=2M/(1-\alpha)$. The uncharged limit of the sparsity parameter is displayed in Fig.~\ref{fig:eta_tsallis_uncharged}. In this case, the electric contribution is absent, so the behavior of $\eta(\tilde S)$ is governed only by the cloud-of-strings parameter $\alpha$, the Lorentz-violating parameter $\ell$, the pressure $P$, and the Tsallis non-extensivity parameter $\delta$. This makes the plot especially useful for isolating the role of the non-extensive deformation in a simpler thermodynamic setting.

Figure~\ref{fig:eta_tsallis_uncharged} shows that the uncharged case is qualitatively simpler than the charged one. Once the term proportional to $q^2$ is removed from Eq.~(\ref{sparsity-2}), the denominator no longer develops the same sharp structures responsible for the large peaks observed in Fig.~\ref{fig:eta_tsallis}. As a consequence, the sparsity parameter varies smoothly with $\tilde S$ throughout the interval considered, providing a cleaner view of the effect of the Tsallis deformation itself.

Another relevant feature is that the ordering of the curves with respect to $\delta$ is preserved over a broad entropy range. This indicates that the non-extensive parameter serves as a robust control parameter for radiation sparsity, even in the neutral black-hole sector. In particular, the separation among the profiles shows that the Hawking cascade remains highly sensitive to the choice of entropy functional, despite the absence of the electric contribution.

From a physical perspective, the neutral configuration isolates the combined action of the bumblebee background, the cloud of strings, and the Tsallis correction without the additional distortion introduced by charge. Therefore, Fig.~\ref{fig:eta_tsallis_uncharged} confirms that non-extensive effects on the sparsity parameter are not merely a charged-sector phenomenon. Rather, they are intrinsic to the modified entropy framework itself and persist even in the simpler uncharged AdS black-hole configuration.

This result is consistent with Eq.~(\ref{sparsity-5}), which shows explicitly that, in the neutral limit, the sparsity parameter continues to depend on $\delta$, $\ell$, and $\alpha$. Hence, even without electric charge, the Hawking emission remains sensitive to Lorentz violation, to the surrounding cloud-of-strings matter distribution, and to the choice of non-extensive entropy model.

From the above discussion, we observe that the sparsity parameter (which is dimensionless) depends not only on the Lorentz-violating parameter $\ell$ and the string cloud parameter $\alpha$ but also on the Tsallis parameter $\delta$, which lies in the range $0 < \delta \leq 3/2$. Thus, this Tsallis parameter modifies the sparsity parameter relative to the Bekestein-Hawking entropy-based result.

\section{Thermodynamics via Barrow entropy-based framework}

It is well known that black holes may exhibit intricate, fractal-like structures at small scales due to quantum gravitational effects, potentially increasing their surface area and entropy \cite{Barrow2020}. This suggests that the standard Bekenstein-Hawking entropy, which assumes a smooth horizon, may underestimate the black hole's true entropy. Barrow proposed a generalized entropy-area relation by introducing a parameter $\epsilon$ that characterizes the fractal structure of the horizon. In this framework, the black hole entropy scales as $S \propto A^{1+\epsilon/2}_{\rm BH}$, indicating that the horizon surface grows faster than in the classical case. This approach captures the small-scale quantum corrections and provides a more general description of black hole thermodynamics, particularly when additional structures such as string clouds or other fields modify the spacetime geometry.

According to the Barrow entropy model \cite{Barrow2020}, the corrected entropy takes the form
\begin{equation}
    S_{\epsilon} \equiv \tilde{S}=(S_{\rm BH})^{1+\epsilon/2}.
\end{equation}
Here $\epsilon$ denotes the Barrow parameter (sometimes called fractal correction parameter), satisfying values ranging in $0 \leq \epsilon \leq 1$, which characterizes the deviation from the standard Bekenstein-Hawking entropy. 

Following a procedure similar to the one used earlier, it can be shown that the thermodynamic properties and the sparsity parameter remain unchanged from those obtained in the Tsallis entropy framework if the deviation parameter is set to $\delta = 1 + \epsilon/2$. Equivalently, the thermodynamic results can be obtained by replacing $\delta$ with $1 + \epsilon/2$ in all expressions for the thermodynamic quantities, including the sparsity parameter.

\section{Conclusions}\label{sec:10}

In this work, we analyzed the thermodynamic properties of charged AdS black holes surrounded by a cloud of strings in bumblebee gravity. The system is controlled by two deformation parameters: the cloud-of-strings parameter $\alpha$, which modifies the matter content surrounding the black hole, and the Lorentz-violating parameter $\ell$, which emerges from the bumblebee sector and changes the effective charge coupling and radial geometry. Working in the extended phase-space framework, we derived the Hawking temperature, entropy, thermodynamic volume, free energies, specific heat, and the corresponding equation of state.

Our results show that both $\alpha$ and $\ell$ have clear and complementary thermodynamic effects. The Hawking temperature is globally suppressed as $\ell$ increases, while larger values of $\alpha$ also reduce the thermal scale by weakening the effective contribution of the constant term in the metric function. The specific heat exhibits Davies-type divergences, indicating second-order transition points between locally unstable and stable branches. At the level of global stability, the Gibbs free energy develops the standard swallowtail structure associated with a first-order small--large black-hole transition, but its location and shape are significantly altered by the Lorentz-violating background and the cloud of strings.

The criticality analysis revealed that the equation of state preserves the Van der Waals-like structure, with a well-defined critical point $(v_c,T_c,P_c)$. However, the universal critical ratio is no longer the standard RN-AdS value $3/8$. Instead, we found
\[
\rho_c=\frac{P_c v_c}{T_c}=\frac{3\sqrt{1+\ell}}{8},
\]
which depends explicitly on the Lorentz-violating parameter and is independent of the string-cloud parameter. This provides a clean thermodynamic signature of Lorentz symmetry breaking in the critical sector. The accompanying plots show that the critical volume, temperature, and pressure respond differently to $\ell$, highlighting the nontrivial deformation of the critical point.

We also discussed the Joule--Thomson expansion in the present framework, deriving the inversion and isenthalpic curves from the thermodynamic relations adopted in this work. The corresponding plots show that both $\ell$ and $\alpha$ shift the inversion locus, thereby modifying the boundary between heating and cooling regions. The isenthalpic curves confirm that the inversion line passes through the maxima of the constant-enthalpy trajectories, as expected in the black-hole analog of ordinary gas expansion.

Finally, we extended the analysis to a Tsallis entropy-based framework. In this non-extensive description, the parameter $\delta$ modifies the scaling of temperature and specific heat with the horizon radius, thereby shifting the stability windows and critical behavior. In particular, the critical specific volume grows with $\delta$ and diverges as $\delta\to 1/2$, indicating that the standard Van der Waals-like critical structure breaks down near that limit. The numerical results further show that Lorentz violation and the cloud of strings remain relevant even in the Tsallis framework, modifying the extent of the non-extensive effects.

Overall, the combined action of the string cloud, the Lorentz-violating bumblebee sector, and the Tsallis entropy deformation yields a rich thermodynamic structure that extends the standard charged AdS black-hole picture. The model therefore provides a useful setting for investigating how symmetry breaking, surrounding matter distributions, and non-extensive entropy can jointly reshape black-hole phase transitions and thermal response.

\section*{Acknowledgments}

F.A. acknowledges the Inter University Center for Astronomy and Astrophysics (IUCAA), Pune, India for granting visiting associateship. E. O. Silva acknowledges the support from Conselho Nacional de Desenvolvimento Cient\'{i}fico e Tecnol\'{o}gico (CNPq) (grants 306308/2022-3), Funda\c c\~ao de Amparo \`{a} Pesquisa e ao Desenvolvimento Cient\'{i}fico e Tecnol\'{o}gico do Maranh\~ao (FAPEMA) (grants UNIVERSAL-06395/22), and Coordena\c c\~ao de Aperfei\c coamento de Pessoal de N\'{i}vel Superior (CAPES) - Brazil (Code 001).

\section*{Data Availability Statement}
There are no new data associated with this article.

\section*{Code/Software}

No code/software were developed in this article.

\end{document}